\documentclass[aps,preprint,groupedaddress,showpacs,floatfix]{revtex4}
\usepackage{graphicx}
\begin{document}

\title{Analytical $\pi\pi$ scattering amplitude and the light scalars-II}
\author {
N.N. Achasov$^{\,a}$ \email{achasov@math.nsc.ru} and A.V.
Kiselev$^{\,a,b}$ \email{kiselev@math.nsc.ru}}

\affiliation{
   $^a$Laboratory of Theoretical Physics,
 Sobolev Institute for Mathematics, 630090, Novosibirsk, Russia\\
$^b$Novosibirsk State University, 630090, Novosibirsk, Russia}

\date{\today}

\begin{abstract}

In the paper Phys. Rev. {\bf D83}, 054008 (2011) we constructed
the $\pi\pi$ scattering amplitude $T^0_0$ with regular analytical
properties in the $s$ complex plane, describing both experimental
data and the results based on chiral expansion and Roy equations.
Now the results obtained during development of our work are
presented. We dwell on questions dealing with the low $\sigma-f_0$
mixing, inelasticity description and the kaon loop model for
$\phi\to \gamma(\sigma+f_0)$ reaction, and show a number of new
fits. In particular, we show that the minimization of the
$\sigma-f_0$ mixing results in the four-quark scenario for light
scalars: the $\sigma$(600) coupling with the $K\bar K$ channel is
suppressed relatively to the coupling with the $\pi\pi$ channel,
and the $f_0$(980) coupling with the $\pi\pi$ channel is
suppressed relatively to the coupling with the $K\bar K$ channel.

The correct analytical properties of the $\pi\pi$ scattering
amplitude are reached with the help of rather complicated
background function. We also suggest much more simple background
parameterization, practically preserving the resonance features,
which is comfortable for experimental data analysis, but allows to
describe the results based on chiral expansion and Roy equations
only on the real $s$ axis.

\end{abstract}


\maketitle

\section{Introduction}

 Study of light scalar resonances is one of the central
problems of nonperturbative QCD, it is important for understanding
the chiral symmetry realization way  resulting from the
confinement physics.

In Refs. \cite{our_f0} we described the high-statistical KLOE data
on the $\phi\to\pi^0\pi^0\gamma$ decay \cite{pi0publ} in the frame
of the kaon loop model $\phi\to
K^+K^-\to\gamma(f_0+\sigma)\to\gamma\pi^0\pi^0$
\cite{achasov-89,achasov-97,a0f0,achasov-03,our_a0} simultaneously
with the data on the $\pi\pi$ scattering and the $\pi\pi\to K\bar
K$ reaction. The description was carried out taking into account
the chiral shielding of the $\sigma (600)$ meson
\cite{annshgn-94,annshgn-07} and its mixing with the $f_0(980)$
meson, the data yielded evidence in favor of the four-quark nature
of the $\sigma (600)$ and $f_0(980)$ mesons.

At the same time it was calculated in Ref. \cite{sigmaPole} the
$\pi\pi$ scattering amplitude in the $s$ complex plane, basing on
chiral expansion, dispersion relations, and Roy equations. In
particular, the pole was obtained at
$s=M_\sigma^2=(6.2-12.3i)\,m_\pi^2$, where

\begin{equation}
M_\sigma = 441^{+16}_{-8} - i272^{+9}_{-12.5}\ \mbox{MeV}\,,
\label{poleSigma}\end{equation}

\noindent which was assigned to the $\sigma$ resonance. Aiming the
comparison of the results of Refs. \cite{our_f0} and
\cite{sigmaPole}, we built up the S-wave $\pi\pi$ scattering
amplitude $T^0_0$ with $I=0$ with correct analytical properties in
the complex $s$ plane \cite{our_f0_2011}. Remain that in our model
the $S$ matrix of the $\pi\pi$ scattering is the product of the
"resonance" and "elastic background" parts:
\begin{equation}
S^0_0 = S^{0\,back}_0\,S^{0\,res}_0\,, \label{SSrepres}
\end{equation}

\noindent and we introduced the special $S^{0\,back}_0$
parametrization to obtain the correct $T^0_0$ analytical
properties ($S^{0\,res}_0$ had correct analytical properties in
Refs. \cite{our_f0} already). In Ref. \cite{our_f0_2011} we
successfully described the experimental data and the Ref.
\cite{sigmaPole} results on the real $s$ axis using the
constructed $\pi\pi$ amplitude, while the $\sigma$ pole was
located rather far from the Ref. \cite{sigmaPole} result. We
assumed that this deviation is caused by approximate character of
the Roy equations, that take into account only the $\pi\pi$ decay
channel. This question will be discussed below in more details.

In this paper we present the enlarged data analysis. We dwell on
the minimization of the $\sigma-f_0$ mixing, that leads to the
four-quark scenario for light scalars: the $\sigma$(600) coupling
with the $K\bar K$ channel is suppressed relatively to the
coupling with the $\pi\pi$ channel, and the $f_0$(980) coupling
with the $\pi\pi$ channel is suppressed relatively to the coupling
with the $K\bar K$ channel \cite{jaffe}. Inelasticity is also
crucial for the analysis, here we describe the peculiar behavior
of the data up to $1.2$ GeV.

In Refs. \cite{our_f0, our_f0_2011} we used the factor $P_K$,
caused by the elastic $K\bar K$ background phase, that allows to
correct the kaon loop model, suggested in Ref. \cite{achasov-89},
under the $K\bar K$ threshold. Now we investigate how small this
correction may be.

The set of new fits and tables is presented in Sec.
\ref{sDataAnalysis}. The residues of the $\pi\pi$ scattering
amplitude and its resonance part in resonance poles are presented
for the first time.

As the analytical background $S^{0\,back}_0$ is a rather
complicated function, in Sec. \ref{sSimpleBack} we suggest much
more simple background parameterization, practically preserving
the resonance features, which is comfortable for experimental data
analysis, though allows to describe the results of Ref.
\cite{sigmaPole} only on the real $s$ axis.

The conclusion is in Sec. \ref{sConclusion}.

Note that the $S^{0\,res}_0$ parameterization and the complicated
background parameterization are the same as in Ref.
\cite{our_f0_2011}. The modification of the $K \bar K$ background
phase is described in Sec. \ref{sDataAnalysis}.

\section{Data analysis, background with the correct analytical properties ("complicated" background)}
\label{sDataAnalysis}

The measure of the $\sigma-f_0$ mixing intensity is the deviation
from the ideal picture, when the $\pi\pi$ scattering phase
$\delta^0_0$ is equal to $90^\circ$ at the $\sigma(600)$ mass
$m_\sigma$, and equal to $270^\circ$ at the $f_0(980)$ mass
$m_{f_0}$. We require these phases, $\delta^0_0 (m_\sigma)$ and
$\delta^0_0 (m_{f_0})$, to be close to their "ideal" values.

Remain that the background phase of the $K\bar K$ scattering,
$\delta_B^{K\bar K}$, changes the modulus of the $K\bar
K\to\pi^0\pi^0$ amplitude under the $K\bar K$ threshold, at $m<2
m_{K}$, in the amplitude $\phi\to K\bar K\to \pi^0\pi^0\gamma$
\cite{Msig}. In Ref. \cite{our_f0_2011} we define

\[P_K= \left\{\begin{array}{ll}
  e^{i\delta_B^{K\bar K}}\hspace{60 mm} m\geq 2
m_{K}\,;\\
 \mbox{analytical continuation of } e^{i\delta_B^{K\bar
 K}}\hspace{12 mm}
m<2 m_{K}\,.\hspace{53 mm}\addtocounter{equation}{1}
(\theequation)
 \label{Kphas}
\end{array}\right.\]

In the present paper we investigate the influence of $P_K$ on the
$\phi\to (f_0+\sigma)\gamma$ amplitude in the $f_0(980)$ region,
$m>850$ MeV. We upgrade the parametrization of the $\delta ^{K
\bar K} (m)$, used in Refs. \cite{our_f0, our_f0_2011}, now the
$\delta_B^{K \bar K}$ is parametrized in the following way:
$$e^{2i\delta_B^{K\bar K}}=\frac{1+i2p_K f_K(m^2)}{1-i2p_K
f_K(m^2)}\,,\ \ p_K=\frac{1}{2}\sqrt{m^2-4m_{K^+}^2}\,,$$
\begin{equation}
f_K(m^2)=-\bigg(1-w+w\frac{(m-m_2)^2/\Lambda_2^2}{1+(m-m_2)^2/\Lambda_2^2}\bigg)\frac{\arctan
(\frac{m^2-m_1^2}{\Lambda_1^2})- \phi_0}{\Lambda _K}\,. \label{fK}
\end{equation} Note that the $P_K$ also provides pole absence in the analytical
continuation of the $\phi\to (f_0+\sigma)\gamma$ amplitude under
the $\pi\pi$ threshold, see Ref. \cite{our_f0}.

The experimental data on the inelasticity $\eta^0_0$, see Fig. 4,
favor the low value near $1.01$ GeV and sharp growth up to $1.2$
GeV. Below it is shown that it is possible to reach such a
behaviour.

Our results for Fits 1-5 are shown in Tables I-VI and Figs. 1-24.
Fits 1-5 show that the allowed range of $\sigma(600)$ and
$f_0(980)$ parameters is rather wide. For example,
$g_{f_0K^+K^-}^2/4\pi$ is $1$ GeV$^2$ in Fit 1 and more than $4$
GeV$^2$ in Fit 5. This result may be important for coordination of
the $g_{f_0K^+K^-}^2/4\pi$ and $g_{a_0K^+K^-}^2/4\pi$
\cite{equalConstants}.

Note that in Fit 4 the $\sigma(600)$ and $f_0(980)$ are coupled
only with the $\pi\pi$ channel and the $K\bar K$ channel
($x_{f_0}=x_\sigma=0$). As seen from Table I and Figs. 9-16, Fit 4
is in excellent agreement with the data and the \cite{sigmaPole}
results.

We introduce 56 parameters, but for restrictions (expresses 5
parameters through others) and parameters (or their combinations),
that go to the bound of the permitted range (9 effective links),
the effective number of free parameters is reduced to 42. But it
is significant that fits describe as the experimental data (65
points), as well as the $\pi\pi$ amplitude from the
\cite{sigmaPole} in the range $-5m_\pi^2 < s < 0.64$ GeV$^2$ which
is treated along with experimental data.

As in \cite{our_f0_2011} we show resonance poles of the $T^0_0$ on
some unphysical sheets of its Riemannian surface, depending on
sheets of the polarization operators $\Pi_R^{ab}(s)$. For this
choice of sheets the imaginary parts of pole positions $M_R$ would
be connected to the full widths of the resonances
($2\mbox{Im}M_R=\Gamma_R=\sum_{ab}\Gamma (R\to ab)$) in case of
metastable states, decaying to several channels. Note also that we
do not show poles for Fit 2.

One can see that the obtained $\sigma (600)$ pole positions lie
far from Eq. (\ref{poleSigma}). In Ref. \cite{our_f0_2011} we
noted, that one of the possible reasons is the approximate
character of the Roy equations, that take into account only the
$\pi\pi$ channel.

\newpage

\vspace{-15pt}\noindent Table I. Properties of the resonances and
main characteristics are shown.

\noindent The resonance masses $m_R$ and widths $\Gamma_R(m_R)$
(which may be called "Breit-Wigner" masses and widths) are
parameters in the resonance propagators, see Ref.
\cite{our_f0_2011}. They have clear physical meaning in contrast
to the resonance poles in the complex plane. \vspace{2pt}
\begin{center}
\begin{tabular}{|c|c|c|c|c|c|}\hline

Fit & 1 & 2 & 3 & 4 & 5 \\ \hline

$m_{f_0}$, MeV & $978.30$ & $974.78$ & $981.49$ & $979.85$ &
$980.40$ \\ \hline

$g_{f_0K^+K^-}$, GeV  & $3.54$ & $4.34$ & $5.01$ & $5.01$ & $7.33$
\\ \hline

$g_{f_0K^+K^-}^2/4\pi$, GeV$^2$  & $1$ & $1.5$ & $2$ & $2$ &
$4.2782$
\\ \hline

$g_{f_0 \pi^+\pi^-}$, GeV  & $-1.3924$ & $-1.6150$ & $-1.9836$ &
$-1.6455$ & $-2.5874$
\\ \hline

$g_{f_0\pi^+ \pi^-}^2/4\pi$, GeV$^2$ & $0.154$ & $0.208$ & $0.313$
& $0.215$ & $0.533$
\\ \hline

$x_{f_0}$ & $0.6367$ & $0.6039$ & $1.1701$ & $0$ & $1.1972$
\\ \hline

$\Gamma_{f_0}(m_{f_0})$, MeV & $56.7$ & $76.6$ & $114.8$ & $79.1$
& $195.5$
\\ \hline

$m_{\sigma}$, MeV & $479.40$ & $471.89$ & $470.87$ & $472.87$ &
$469.94$ \\ \hline

$g_{\sigma\pi^+ \pi^-}$, GeV & $2.6676$ & $2.6614$ & $2.7190$ &
$2.7093$ & $2.7362$ \\ \hline

$g_{\sigma\pi^+ \pi^-}^2/4\pi$, GeV$^2$ & $0.564$ & $0.569$ & $
0.588$ & $0.584$ & $0.596$ \\ \hline

$g_{\sigma K^+K^-}$, GeV & $0.553$ & $0.101$ & $0.279$ & $0.274$ &
$0.149$
\\ \hline

$g_{\sigma K^+K^-}^2/4\pi$, GeV$^2$ & $0.001$ & $0.048$ & $0.006$
& $0.006$ & $0.002$ \\ \hline

$x_\sigma$ & $1.1822$ & $0.9187$ & $1.7336$ & $0$ & $1.6291$
\\ \hline

$\Gamma_{\sigma}(m_{\sigma})$, MeV & $362.1$ & $363.2$ & $379.5$ &
$376.0$ & $384.7$
\\ \hline

$C$, GeV$^2$ & $0.05120$ & $0.04465$ & $0.01307$ & $0.00167$ &
$0.03345$ \\ \hline

$\delta $, $^{\circ}$ & $-64.69$& $-58.7$ & $-64.6$ & $-55.4$ &
$-44.0$ \\ \hline

$a^0_0,\ m_\pi^{-1}$ & $0.223$ & $0.220$ & $0.224$ & $0.223$ &
$0.225$ \\ \hline

Adler zero in $\pi\pi\to\pi\pi $ & ($93.5$ MeV)$^2$ & ($85.6$
MeV)$^2$ & ($96.8$ MeV)$^2$ & ($94.6$ MeV)$^2$ & ($92.3$ MeV)$^2$
\\ \hline

$\delta_0^{0\,res}(m_\sigma)$, $^{\circ}$ & $91.8$& $94.1$ &
$91.0$ & $90.6$ & $92.3$
\\ \hline

$\delta_0^{0\,res}(m_{f_0})$, $^{\circ}$ & $250.1$& $250.1$ &
$260.1$ & $255.1$ & $258.7$
\\ \hline

$\eta^0_0$($1010$ MeV) & $0.55$ & $0.52$ & $0.51$ & $0.51$ &
$0.51$ \\ \hline

$\chi^2_{phase}$ ($44$ points) & $53.1$ & $48.9$ & $42.0$ & $40.0$
& $55.1$
\\ \hline

$\chi^2_{sp}$ ($18$ points) & $21.2$ & $20.8$ & $21.3$ & $17.0$ &
$12.6$
\\ \hline
\end{tabular}
\end{center}

\newpage

\begin{center}
\vspace{-15pt}Table II. Parameters of the $K\bar K$ background
phase, $\delta_B^{K\bar K}$, are shown. \vspace{2pt}

\begin{tabular}{|c|c|c|c|c|c|}\hline

Fit & 1 & 2 & 3 & 4 & 5 \\ \hline

$\Lambda _K$, GeV & $0.975$ & $1.245$ & $1.375$ & $1.450$ &
$1.894$ \\ \hline

$\Lambda_1$, MeV & $381.56$ & $ 404.49$ & $387.56$ & $412.43$ &
$322.93$ \\ \hline

$\Lambda _2$, MeV & $83.113$ & $81.137$ & $86.246$ & $65.000$ &
$68.041$ \\ \hline

$m_1$, MeV & $827.48$ & $823.54$ & $801.40$ & $791.48$ & $808.17$
\\ \hline

$m_2$, MeV & $909.17$ & $923.55$ & $911.59$ & $970.52$ & $963.55$
\\ \hline

$w$, MeV & $0.471$ & $0.618$ & $0.492$ & $0.750$ & $0.750$ \\
\hline

$\phi_0$ & $-0.299$ & $0.021$ & $0.153$ & $0.271$ & $0.622$ \\
\hline

\end{tabular}
\end{center}\vspace{10pt}

As it was shown in the SU(2)$\times$SU(2) linear $\sigma$ model,
Ref. \cite{annshgn-07}, the residue of the $\sigma$ pole in the
amplitude of the $\pi\pi$ scattering can not be connected to
coupling constant in the Hermitian (or quasi-Hermitian)
Hamiltonian for it has a large imaginary part. Here we calculate
the residues of the amplitude $T^0_0$ in the $\sigma (600)$ pole,
see Table V, and illustrate this fact in our case. Note that large
imaginary part is both in the residues of the full amplitude
$T^0_0$ and its resonance part $T^{0\,Res}_0$. So, considering the
residue of the $\sigma$ pole in $T^0_0$ or $T^{0\,Res}_0$ as
proportional to the square of its coupling constant to the
$\pi\pi$ channel is not a clear guide to understanding the
$\sigma$ meson nature. In addition, this pole can not be
interpreted as a physical state for its huge width.

One can see from Figs. 8b, 16, and 24a that for Fits 2-5 with
$g_{f_0K^+K^-}^2/4\pi \geq 1.5$ GeV$^2$ the maximum of the
$|P_K|^2$ is close to $1$ (about $1.2$), this means that the
correction to the kaon-loop model \cite{achasov-89} is small. For
lower $g_{f_0K^+K^-}^2/4\pi$ the $|P_K|^2$ increases as a
compensation, see Fit 1 and Fig. 8a. This results in a model
dependence of the constant determination. A precise measurement of
the inelasticity $\eta^0_0$ would resolve this problem.

One can see from Table I that for all Fits 1-5 the resonance phase
$\delta^{res}(m)$ is close to $90^{\circ}$ at $m_\sigma$ and to
$270^{\circ}$ at $m_{f_0}$, see also Figs. 3, 11, and 19a.

\newpage

\begin{center}
\vspace{-15pt}Table III. Parameters of the first background
($P_{\pi 1}$), see Ref. \cite{our_f0_2011}, are shown.
\vspace{2pt}
\begin{tabular}{|c|c|c|c|c|c|}\hline

Fit & 1 & 2 & 3 & 4 & 5 \\ \hline

$a_1$ & $-2.767$ & $-1.997$ & $-2.727$ & $-3.152$ & $-2.320$ \\
\hline

$a_2$ & $0.00997$ & $0.02824$ & $0.01228$ & $0.00995$ & $0.00987$
\\ \hline

$a_3$ & $0$ & $0$ & $0$ & $0$ & $0$ \\ \hline

$a_4$  & $2.4774$ & $1.1655$ & $1.9460$ & $3.6119$ & $1.9579$
\\ \hline

$\alpha_1$, GeV$^2$ & $430.875$ & $-3.472$ & $299.566$ & $187.438$
& $230.647$
\\ \hline

$\alpha_2$, GeV$^4$ & $1038.375$ & $802.006$ & $1006.643$ &
$924.912$ & $876.525$
\\ \hline

$\alpha_3$, GeV$^6$ & $853.500$ & $810.211$ & $840.573$ &
$805.455$ & $805.900$ \\ \hline

$\alpha_4$, GeV$^8$ & $237.251$ & $239.362$ & $232.860$ &
$225.823$ & $233.065$
\\ \hline

$\alpha_5$, GeV$^{10}$ & $25.3514$ & $25.4850$ & $24.8635$ &
$24.9756$ & $25.2960$ \\ \hline

$\alpha_6$, GeV$^{12}$ & $0.248630$ & $ 0.218526$ & $0.225182$ &
$0.240893$ & $0.224103$ \\ \hline

$c_1$, GeV & $504.558$ & $680.672$ & $543.245$ & $499.429$ &
$557.733$
\\ \hline

$c_2$, GeV$^3$ & $-2745.58$ & $-2246.19$ & $-2532.80$ & $-2395.72$
& $-2484.57$
\\ \hline

$c_3$, GeV$^5$ & $132.007$ & $176.850$ & $226.617$ & $256.520$ &
$191.569$
\\ \hline

$c_4$, GeV$^7$ & $390.262$ & $379.216$ & $399.808$ & $404.615$ &
$394.230$
\\ \hline

$c_5$, GeV$^9$ & $50.6689$ & $51.7071$ & $50.4545$ & $49.8151$ &
$51.4728$ \\ \hline

$c_6$, GeV$^{11}$ & $-0.612729$& $-0.636956$ & $-0.633709$ &
$-0.711296$ & $-0.646822$ \\ \hline

$m_1$, MeV & $921.52$ & $766.81$ & $1049.46$ & $1088.06$ &
$915.18$
\\ \hline

$g_1$, MeV & $301.06$ & $302.40$ & $302.30$ & $333.34$ & $330.83$
\\ \hline

$m_2$, MeV & $1395.84$ & $937.07$ & $970.74$ & $1104.06$ &
$1025.77$ \\ \hline

$g_2$, MeV & $367.05$ & $305.90$ & $310.37$ & $421.12$ & $324.93$
\\ \hline

$m_3$, MeV & $1208.42$ & $1432.57$ & $1098.81$ & $1125.51$ &
$1330.56$ \\ \hline

$g_3$, MeV & $335.62$ & $304.46$ & $388.57$ & $378.42$ & $301.39$
\\ \hline

$m_4$, MeV & $1078.40$ & $898.85$ & $1053.21$ & $1162.64$ &
$907.71$ \\ \hline

$g_4$, MeV & $395.12$ & $429.54$ & $403.00$ & $388.93$ & $437.65$
\\ \hline

$m_5$, MeV & $1011.58$ & $991.70$ & $1017.19$ & $1051.17$ &
$1008.35$ \\ \hline

$g_5$, MeV & $ 499.99$ & $503.50$ & $507.03$ & $500.41$ & $502.24$
\\ \hline

$m_6$, MeV & $ 932.07$ & $1240.82$ & $1174.80$ & $1154.23$ &
$1264.43$ \\ \hline

$g_6$, MeV & $535.90$ & $616.91$ & $574.55$ & $542.66$ & $629.02$
\\ \hline

\end{tabular}
\end{center}

\newpage
\begin{center}
%\vspace{-15pt}
\noindent \hspace{-60pt}Table IV. Parameters of the second
background ($P_{\pi 2}$), see Ref. \cite{our_f0_2011}, are shown.
\vspace{2pt}
\begin{tabular}{|c|c|c|c|c|c|}\hline

Fit & 1 & 2 & 3 & 4 & 5 \\ \hline

$\Lambda$, MeV & $88.506$ & $78.187$ & $73.302$ & $70.443$ &
$68.804$ \\ \hline

$k_2$ & $0.0295157$ & $0.0310661$ & $0.0313951$ & $0.0415406$ &
$0.0250125$ \\ \hline

$\beta$ & $143.367$ & $132.832$ & $125.729$ & $126.711$ &
$165.179$ \\ \hline

$\gamma_1$ & $383.759$ & $330.693$ & $365.564$ & $374.814$ &
$450.415$ \\ \hline

$\gamma_2$ & $21.5451$ & $24.7134$ & $22.9274$ & $22.1454$ &
$23.1322$ \\ \hline

$m_{1A}$, MeV & $687.75$ & $668.05$ & $721.92$ & $837.74$ &
$607.47$ \\ \hline

$g_{1A}$, MeV & $301.22$ & $303.07$ & $305.36$ & $392.13$ &
$343.56$ \\ \hline

$m_{2A}$, MeV & $495.44$ & $554.28$ & $492.36$ & $487.79$ &
$515.18$ \\ \hline

$g_{2A}$, MeV & $504.69$ & $488.58$ & $457.53$ & $454.08$ &
$494.25$ \\ \hline

$m_{3A}$, MeV & $631.67$ & $637.64$ & $625.43$ & $599.93$ &
$675.83$ \\ \hline

$g_{3A}$, MeV & $117.86$ & $101.74$ & $101.00$ & $101.35$ &
$100.98$ \\ \hline

\end{tabular}
\end{center}\vspace{10pt}

\noindent \hspace{0pt}Table V. The $\sigma(600)$ poles (in MeV),
the residues of $T^0_0$, Res$\,T^0_0$, and of the resonance part
$T^{0 Res}_0$, Res$\,T^{0 Res}_0$, (in 0.01 GeV$^2$) in this pole
on different sheets of the complex $s$ plane depending on sheets
of polarization operators $\Pi^{ab}(s)$ are shown.
\begin{center}
\vspace{-13pt}
\begin{tabular}{|c|c|c|c|c|c|c|c|c|c|c|}\hline

\multicolumn{5}{|c|}{Sheets of $\Pi^{ab}$} &
\multicolumn{3}{|c|}{Fit 1} & \multicolumn{3}{|c|}{Fit 5}\\ \hline

$\Pi^{\pi\pi}$ & $\Pi^{K\bar K}$ & $\Pi^{\eta\eta}$ &
$\Pi^{\eta\eta'}$ & $\Pi^{\eta'\eta'}$ & $\sigma$ pole &
Res$\,T^0_0$ & Res$\,T^{0 Res}_0$ & $\sigma$ pole & Res$\,T^0_0$ &
Res$\,T^{0 Res}_0$\\ \hline

II & I & I & I & I & $565 - 204\,i$ & $-2+14\,i$ & $-22-11\,i$ &
$566 - 201\,i$ & $-1+13\,i$ & $-20-12\,i$
\\ \hline

II & II & I & I & I & $612 - 346\,i$ & $5+3\,i$ & $-18+14\,i$ &
$569 - 267\,i$ & $3+10\,i$ & $-26-1\,i$
\\ \hline

II & II & II & I & I & $542 - 396\,i$ & $-1+3\,i$ & $-16-3\,i$ &
$522 - 379\,i$ & $-2+3\,i$ & $-14-6\,i$
\\ \hline

II & II & II & II & I & $577 - 522\,i$ & $0.2+1\,i$ & $-15-1\,i$ &
$612-626\,i$ & $0.3+0.4\,i$ & $-15-4\,i$
\\ \hline

II & II & II & II & II & $633 - 534\,i$ & $1+1\,i$ & $-23+2\,i$ &
$664 - 651\,i$ & $0.5+0.3\,i$ & $-19-4\,i$
\\ \hline

\multicolumn{5}{|c|}{Sheets of $\Pi^{ab}$} &
\multicolumn{3}{|c|}{Fit 3} & \multicolumn{3}{|c|}{Fit 4}\\ \hline

$\Pi^{\pi\pi}$ & $\Pi^{K\bar K}$ & $\Pi^{\eta\eta}$ &
$\Pi^{\eta\eta'}$ & $\Pi^{\eta'\eta'}$ & $\sigma$ pole &
Res$\,T^0_0$ & Res$\,T^{0 Res}_0$ & $\sigma$ pole & Res$\,T^0_0$
&Res$\,T^{0 Res}_0$\\ \hline

II & I & I & I & I & $572 - 206\,i$ & $-2+14\,i$ & $-21-12\,i$ &
$579 - 216\,i$ & $-3+15\,i$ & $-23-12\,i$
\\ \hline

II & II & I & I & I & $572 - 279\,i$ & $3+10\,i$ & $-26+2\,i$ &
$579 - 273\,i$ & $1+12\,i$ & $-27-1\,i$
\\ \hline

II & II & II & I & I & $526 - 395\,i$ & $-2+2\,i$ & $-12-5\,i$ &
-- & -- & --
\\ \hline

II & II & II & II & I & $623 - 651\,i$ & $0.3+0.3\,i$ & $-14-3\,i$
& -- & -- & --
\\ \hline

II & II & II & II & II & $683 - 679\,i$ & $1+0.1\,i$ & $-19-4\,i$
& -- & -- & --
\\ \hline

\end{tabular}
\end{center}\vspace{10pt}

\newpage

\begin{figure}[t]
\begin{center}
\begin{tabular}{ccc}
\includegraphics[width=8cm]{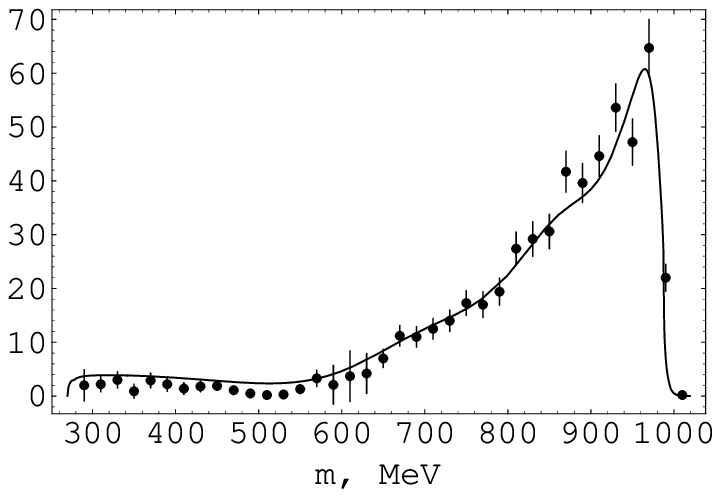}& \includegraphics[width=8cm]{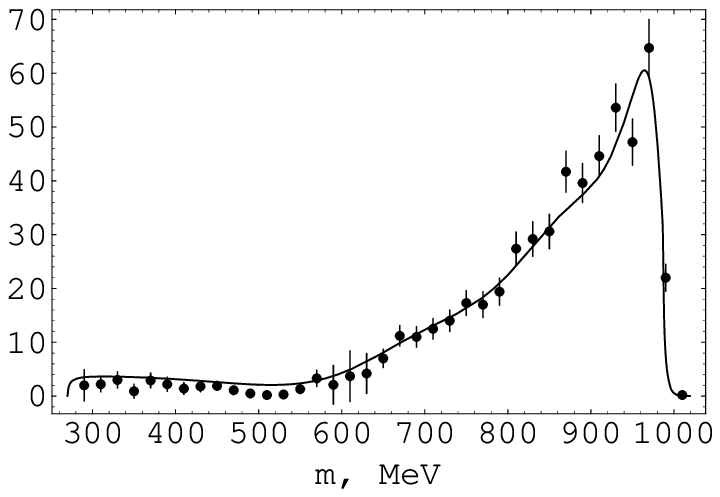}\\ (a)&(b)
\end{tabular}
\end{center}
\caption{The $\pi^0\pi^0$ spectrum in the
$\phi\to\pi^0\pi^0\gamma$ decay, theoretical curve, and the KLOE
data (points) \cite{pi0publ} are shown: a) Fit 1, b) Fit 2.}
\label{fig1}
\end{figure}

\noindent Table VI. The $f_0(980)$ poles (in MeV), the residues of
$T^0_0$, Res$\,T^0_0$, and of the resonance part $T^{0 Res}_0$,
Res$\,T^{0 Res}_0$, (in 0.01 GeV$^2$) in this pole on different
sheets of the complex $s$ plane depending on sheets of
polarization operators $\Pi^{ab}(s)$ are shown.
\begin{center}
\vspace{-13pt}
\begin{tabular}{|c|c|c|c|c|c|c|c|c|c|c|}\hline

\multicolumn{5}{|c|}{Sheets of $\Pi^{ab}$} &
\multicolumn{3}{|c|}{Fit 1} & \multicolumn{3}{|c|}{Fit 5}\\ \hline

$\Pi^{\pi\pi}$ & $\Pi^{K\bar K}$ & $\Pi^{\eta\eta}$ &
$\Pi^{\eta\eta'}$ & $\Pi^{\eta'\eta'}$ & $f_0$ pole & Res$\,T^0_0$
& Res$\,T^{0 Res}_0$ & $f_0$ pole & Res$\,T^0_0$ & Res$\,T^{0
Res}_0$\\ \hline

II & I & I & I & I & $986-26\,i$ & $6-2\,i$ & $-7+2\,i$ & $986 -
21\,i$ & $5-1\,i$ & $-6+0.1\,i$
\\ \hline

II & II & I & I & I & $913 - 302\,i$ & $10+5\,i$ & $-19-19\,i$ &
$1575 - 553\,i$ & $-8-4\,i$ & $-21-23\,i$
\\ \hline

II & II & II & I & I & $966 - 450\,i$ & $3-1\,i$ & $-12-10\,i$ &
$2101 - 1065\,i$ & $0.1+5\,i$ & $-28-10\,i$
\\ \hline

II & II & II & II & I & $962 - 465\,i$ & $3-0.3\,i$ & $-12-12\,i$
& $2173 - 1158\,i$ & $1+5\,i$ & $-25-11\,i$
\\ \hline

II & II & II & II & II & $954 - 586\,i$ & $1+0.4\,i$ & $-3-14\,i$
& $2452 - 1570\,i$ & $3+3\,i$ & $-22-10\,i$
\\ \hline

\multicolumn{5}{|c|}{Sheets of $\Pi^{ab}$} &
\multicolumn{3}{|c|}{Fit 3} & \multicolumn{3}{|c|}{Fit 4}\\ \hline

$\Pi^{\pi\pi}$ & $\Pi^{K\bar K}$ & $\Pi^{\eta\eta}$ &
$\Pi^{\eta\eta'}$ & $\Pi^{\eta'\eta'}$ & $f_0$ pole & Res$\,T^0_0$
& Res$\,T^{0 Res}_0$ & $f_0$ pole & Res$\,T^0_0$ & Res$\,T^{0
Res}_0$\\ \hline

II & I & I & I & I & $986 - 23\,i$ & $6-1\,i$ & $-6+1\,i$ & $985 -
20\,i$ & $5-1\,i$ & $-5+1\,i$
\\ \hline

II & II & I & I & I & $1149 - 485\,i$ & $3-6\,i$ & $-14-16\,i$ &
$1187 - 618\,i$ & $0.5-2\,i$ & $-11-9\,i$
\\ \hline

II & II & II & I & I & $1441 - 835\,i$ & $-3+0.4\,i$ & $-20-7\,i$
& -- & -- & --
\\ \hline

II & II & II & II & I & $1469 - 885\,i$ & $-2+0.4\,i$ & $-16-9\,i$
& -- & -- & --
\\ \hline

II & II & II & II & II & $1607 - 1182\,i$ & $-1+1\,i$ & $-11-8\,i$
& -- & -- & --
\\ \hline

\end{tabular}
\end{center}

\begin{figure}[t]
\begin{center}
\begin{tabular}{ccc}
\includegraphics[width=8cm]{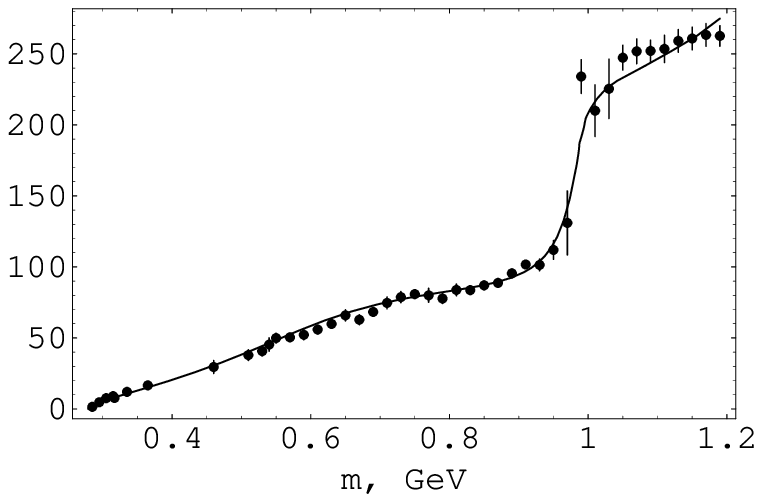}& \raisebox{-1mm}{$\includegraphics[width=8cm]{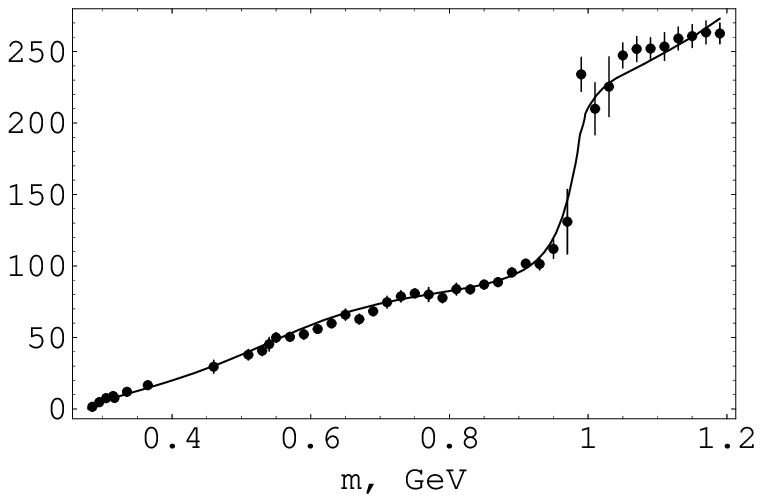}$}\\ (a)&(b)
\end{tabular}
\end{center}
\caption{The phase $\delta_0^0$ of the $\pi\pi$ scattering
(degrees) is shown: a) Fit 1, b) Fit 2. The experimental data from
Refs. \cite{hyams,estabrook,martin,srinivasan,rosselet}.}
\label{fig2}
\end{figure}

\begin{figure}[t]
\begin{center}
\begin{tabular}{ccc}
\includegraphics[width=7cm]{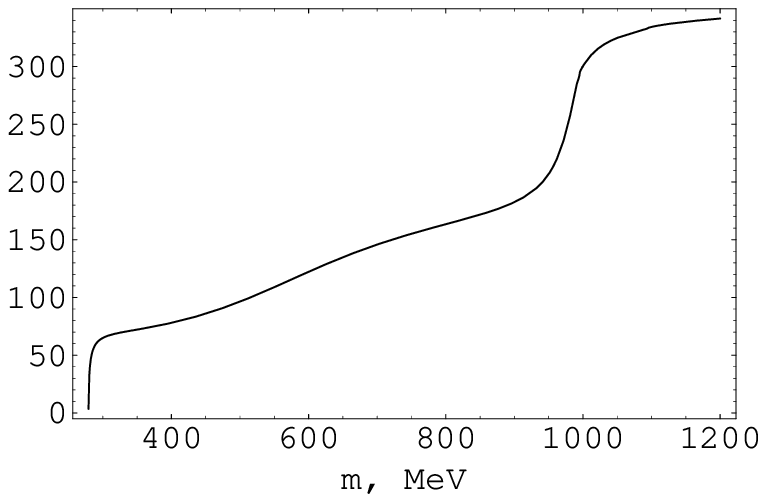}& \raisebox{-1mm}{$\includegraphics[width=7cm]{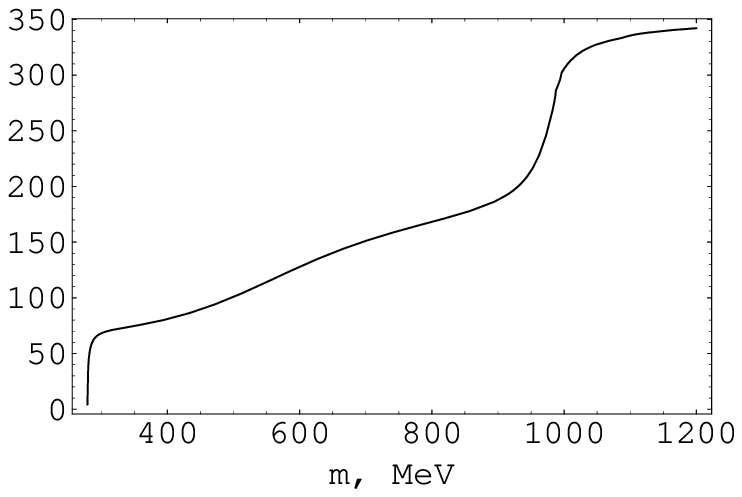}$}\\ (a)&(b)
\end{tabular}
\end{center}
\caption{The resonance phase of the $\pi\pi$ scattering
$\delta_0^{0\,res}$ (degrees) is shown: a) Fit 1, b) Fit 2.}
\label{fig3}
\end{figure}

\begin{figure}[t]
\begin{center}
\begin{tabular}{ccc}
\includegraphics[width=8cm]{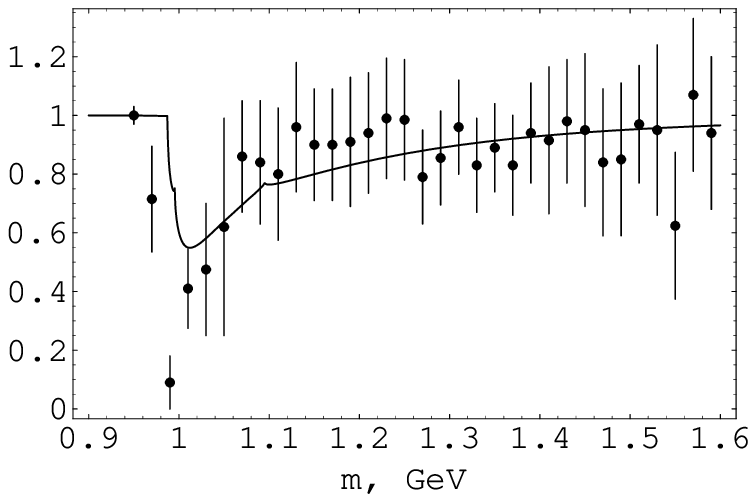}& \raisebox{-1mm}{$\includegraphics[width=8cm]{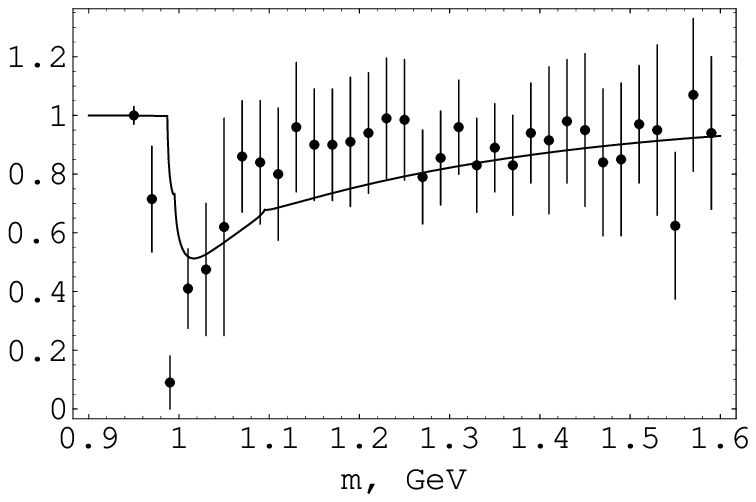}$}\\ (a)&(b)
\end{tabular}
\end{center}
\caption{The inelasticity $\eta^0_0$ is shown: a) Fit 1, b) Fit
2.} \label{fig4}
\end{figure}

\begin{figure}[p]
\begin{center}
\begin{tabular}{ccc}
\includegraphics[width=8cm]{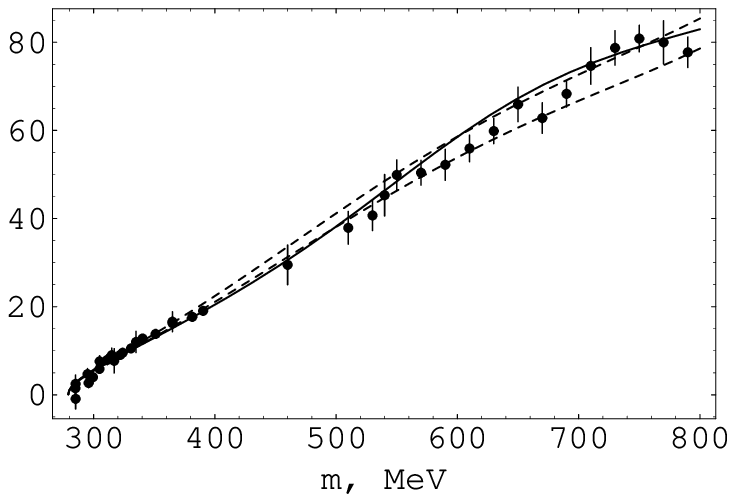}& \raisebox{-1mm}{$\includegraphics[width=8cm]{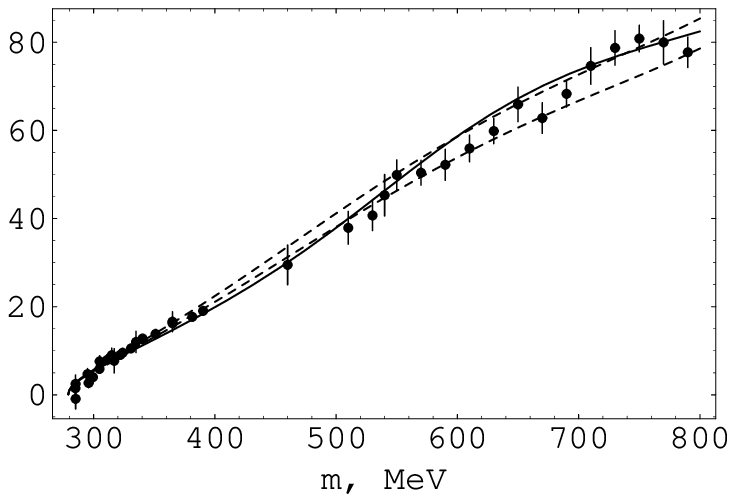}$}\\ (a)&(b)
\end{tabular}
\end{center}
\caption{The phase $\delta_0^0$ of the $\pi\pi$ scattering is
shown. The solid line is our description, dashed lines mark
borders of the corridor \cite{sigmaPole}, and points are the
experimental data from Refs. \cite{scatbnl,na48,
hyams,estabrook,martin,srinivasan,rosselet}: a) Fit 1, b) Fit 2.}
\label{fig5}
\end{figure}

\begin{figure}[p]
\begin{center}
\begin{tabular}{ccc}
\includegraphics[width=8cm]{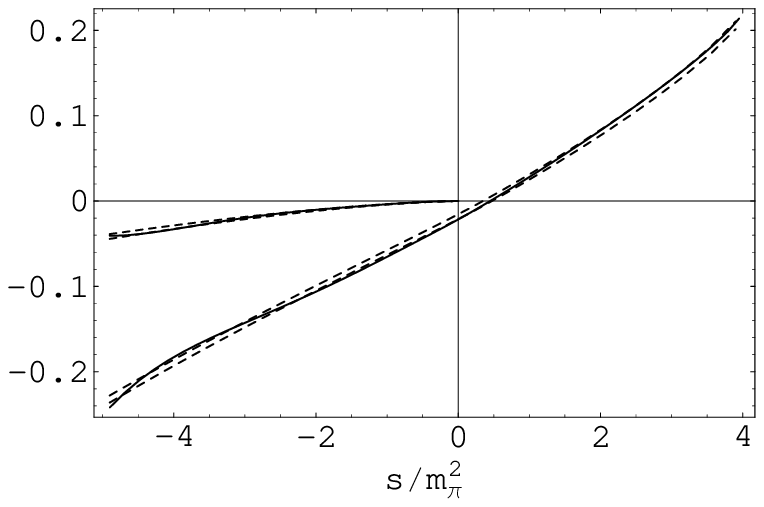}& \raisebox{-1mm}{$\includegraphics[width=8cm]{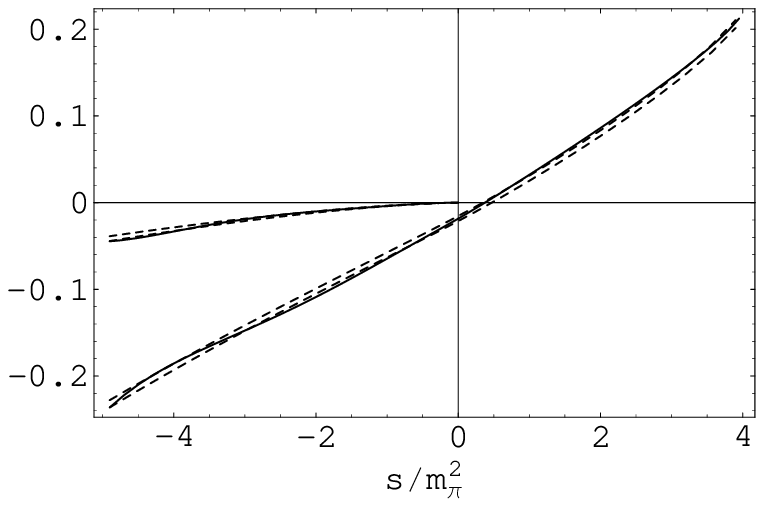}$}\\ (a)&(b)
\end{tabular}
\end{center}
\caption{The real and the imaginary parts of the amplitude $T^0_0$
of the $\pi\pi$ scattering are shown. Solid lines show our
description, dashed lines mark borders of the real part corridor
and the imaginary part for $s < 0$ \cite{sigmaPole}: a) Fit 1; b)
Fit 2. } \label{fig6}
\end{figure}

\begin{figure}[p]
\begin{center}
\begin{tabular}{ccc}
\includegraphics[width=8cm]{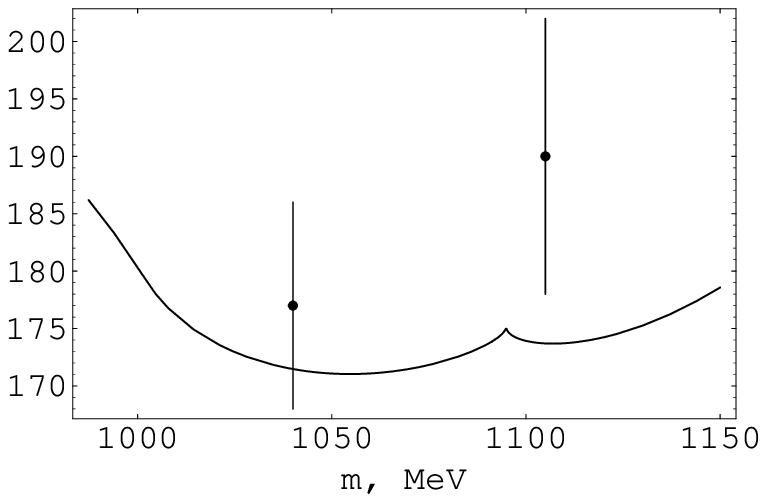}& \raisebox{-1mm}{$\includegraphics[width=8cm]{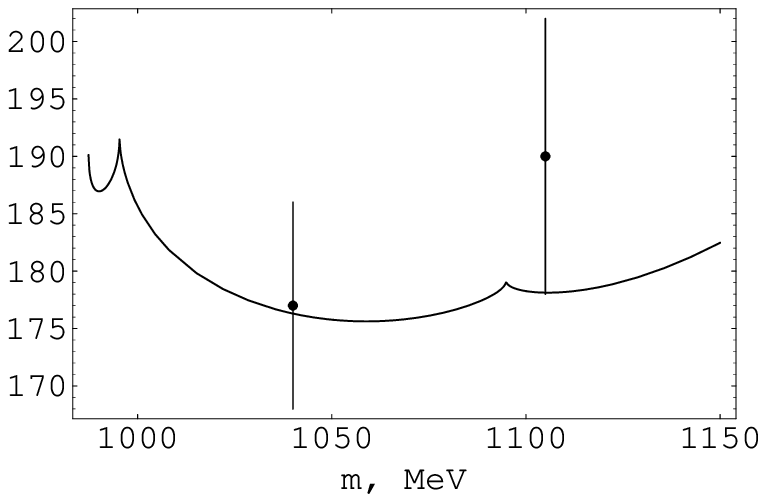}$}\\ (a)&(b)
\end{tabular}
\end{center}
\caption{The phase $\delta^{\pi K}$ of the $\pi\pi\to K\bar K $
scattering is shown: a) Fit 1; b) Fit 2.} \label{fig7}
\end{figure}

\begin{figure}[p]
\begin{center}
\begin{tabular}{ccc}
\includegraphics[width=8cm]{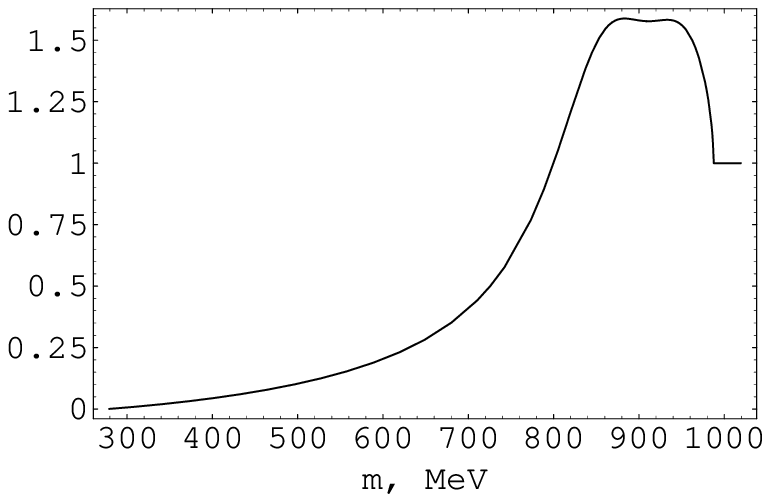}& \raisebox{-1mm}{$\includegraphics[width=8cm]{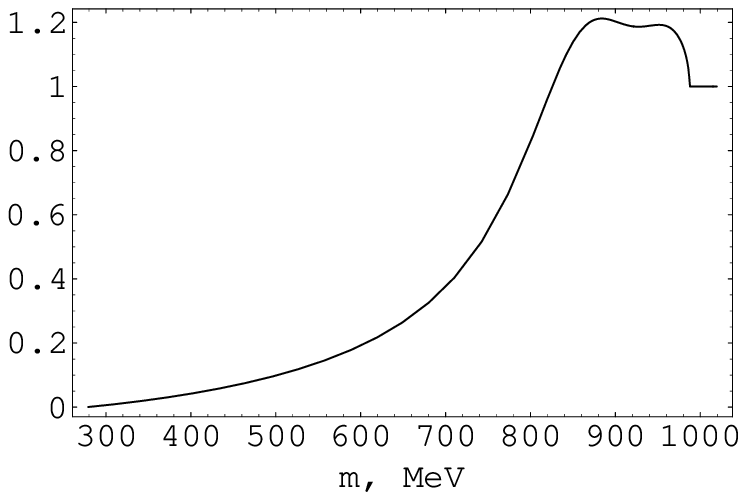}$}\\ (a)&(b)
\end{tabular}
\end{center}
\caption{The $|P_K(m)|^2$ is shown, see Eq. (7): a) Fit 1; b) Fit
2.} \label{fig8}
\end{figure}

\begin{figure}[p]
\begin{center}
\begin{tabular}{ccc}
\includegraphics[width=8cm]{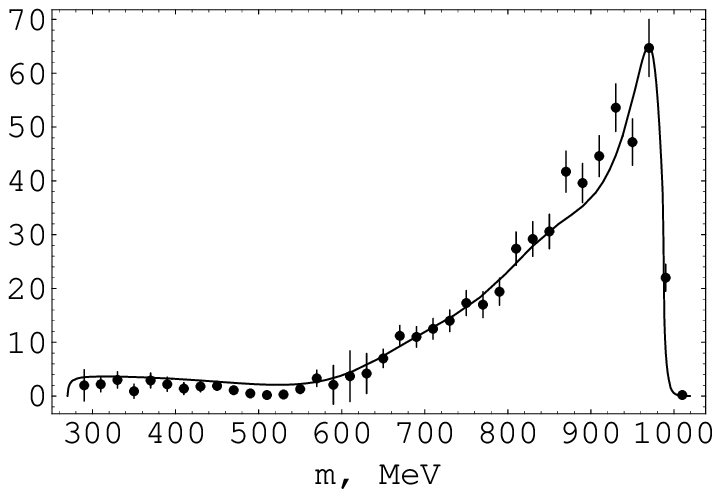}& \includegraphics[width=8cm]{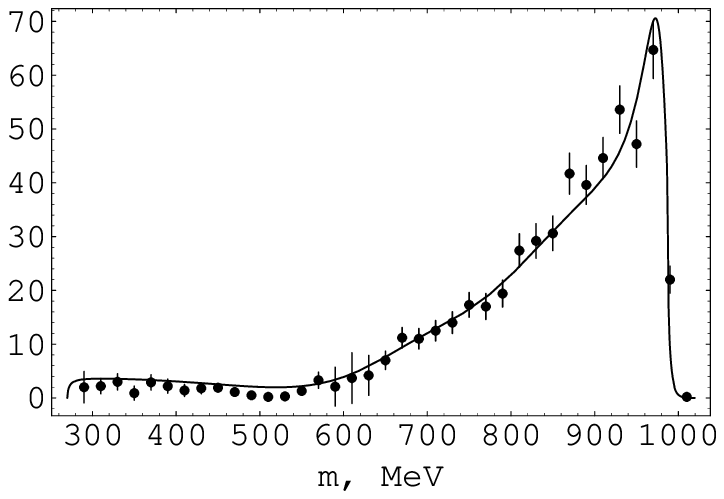}\\ (a)&(b)
\end{tabular}
\end{center}
\caption{The $\pi^0\pi^0$ spectrum in the
$\phi\to\pi^0\pi^0\gamma$ decay, theoretical curve, and the KLOE
data (points) \cite{pi0publ} are shown: a) Fit 3, b) Fit 4.}
\label{fig9}
\end{figure}

\begin{figure}[p]
\begin{center}
\begin{tabular}{ccc}
\includegraphics[width=8cm]{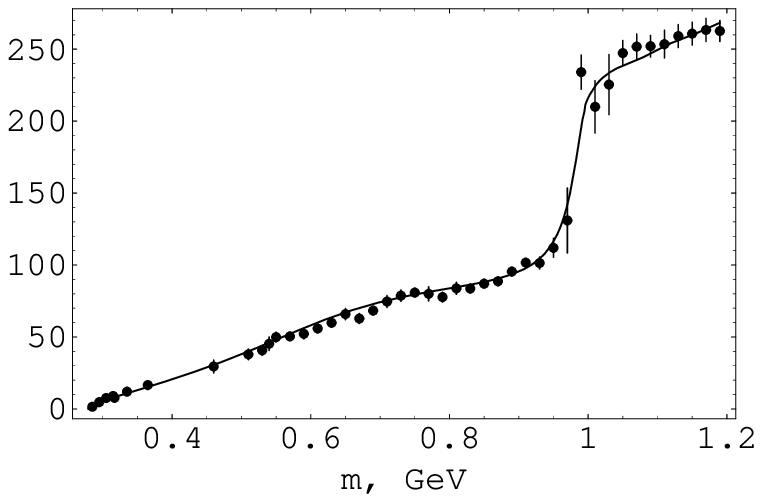}& \raisebox{-1mm}{$\includegraphics[width=8cm]{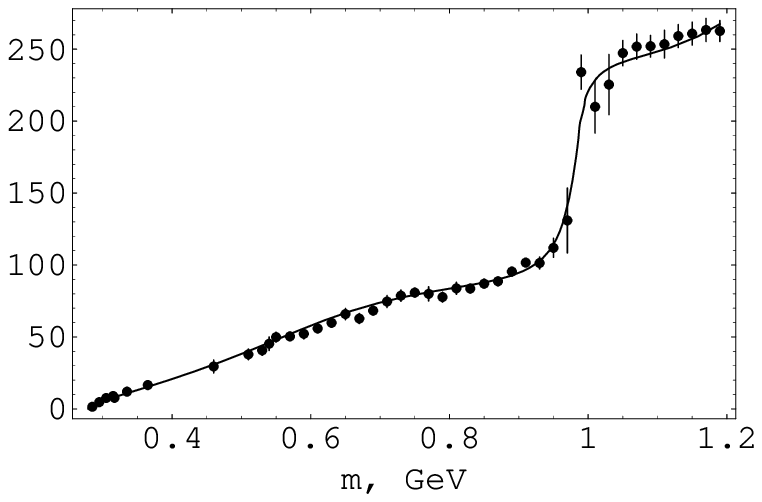}$}\\ (a)&(b)
\end{tabular}
\end{center}
\caption{The phase $\delta_0^0$ of the $\pi\pi$ scattering
(degrees) is shown: a) Fit 3, b) Fit 4. The experimental data from
Refs. \cite{hyams,estabrook,martin,srinivasan,rosselet}.}
\label{fig10}
\end{figure}

\begin{figure}[p]
\begin{center}
\begin{tabular}{ccc}
\includegraphics[width=8cm]{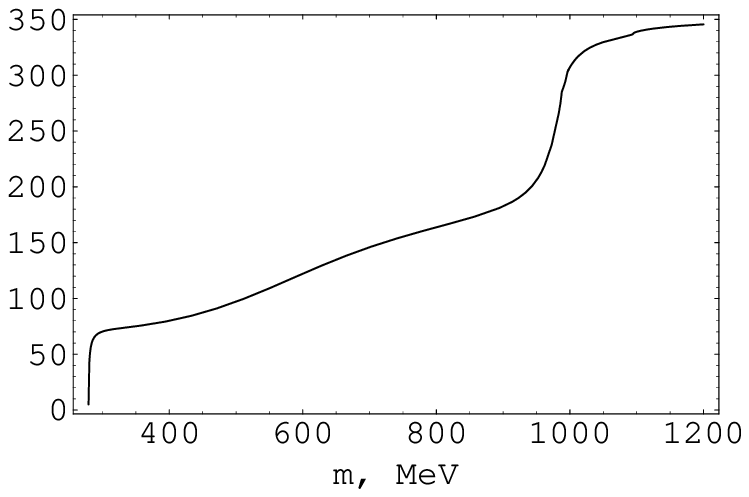}& \raisebox{-1mm}{$\includegraphics[width=8cm]{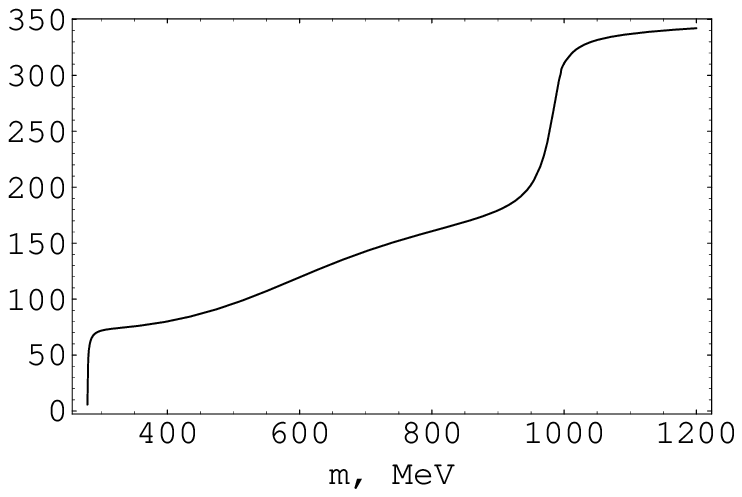}$}\\ (a)&(b)
\end{tabular}
\end{center}
\caption{The resonance phase of the $\pi\pi$ scattering
$\delta_0^{0\,res}$ (degrees) is shown: a) Fit 3, b) Fit 4.}
\label{fig11}
\end{figure}

\begin{figure}[h]
\begin{center}
\begin{tabular}{ccc}
\includegraphics[width=8cm]{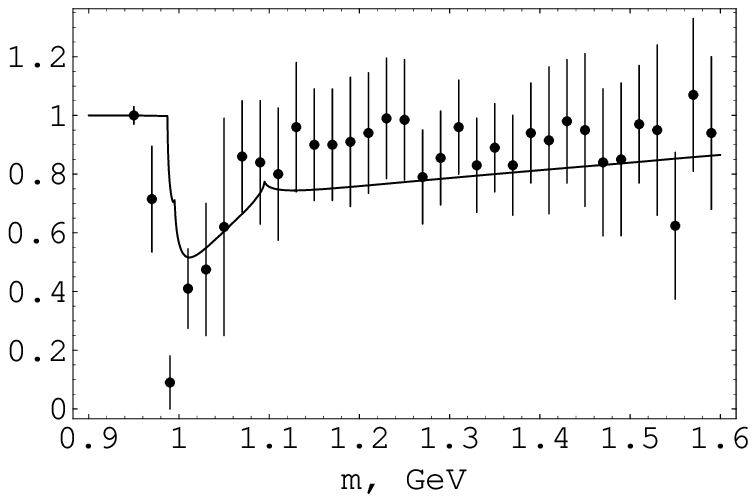}& \raisebox{-1mm}{$\includegraphics[width=8cm]{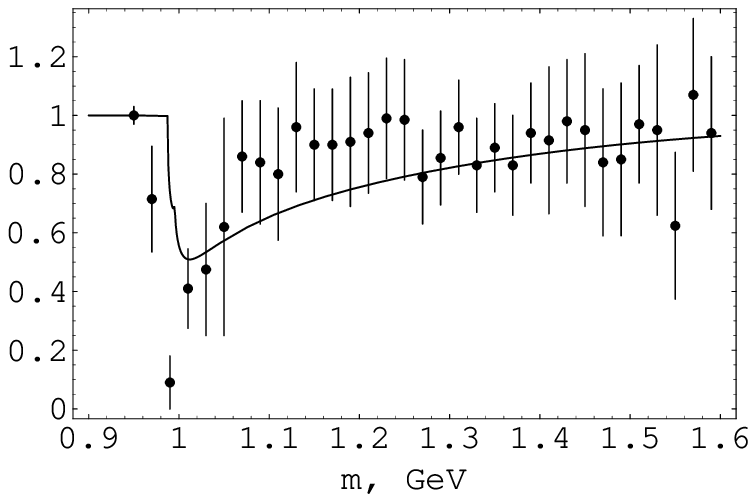}$}\\ (a)&(b)
\end{tabular}
\end{center}
\caption{The inelasticity $\eta^0_0$ is shown: a) Fit 3, b) Fit
4.} \label{fig12}
\end{figure}

\begin{figure}[p]
\begin{center}
\begin{tabular}{ccc}
\includegraphics[width=8cm]{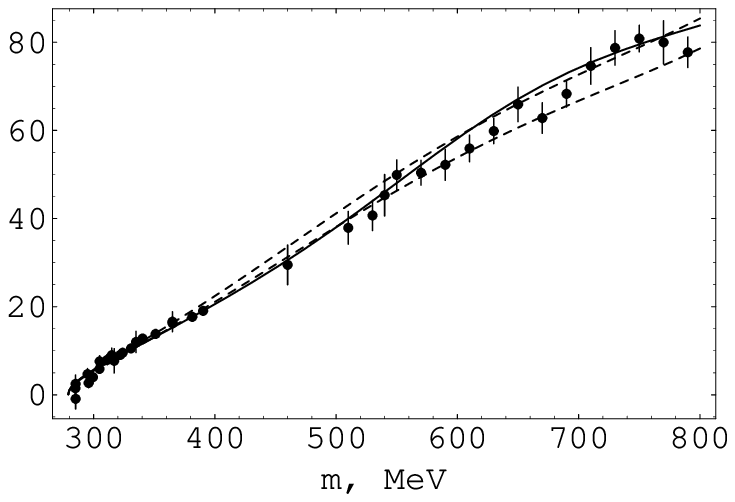}& \raisebox{-1mm}{$\includegraphics[width=8cm]{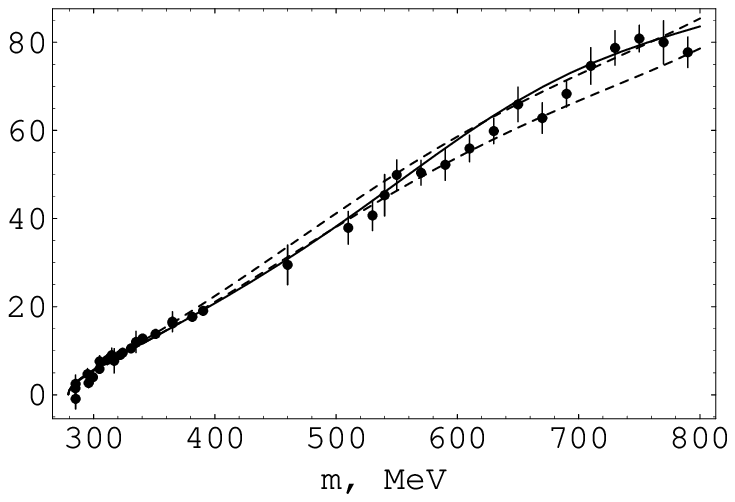}$}\\ (a)&(b)
\end{tabular}
\end{center}
\caption{The phase $\delta_0^0$ of the $\pi\pi$ scattering is
shown. The solid line is our description, dashed lines mark
borders of the corridor \cite{sigmaPole}, and points are the
experimental data from Refs. \cite{scatbnl,na48,
hyams,estabrook,martin,srinivasan,rosselet}: a) Fit 3, b) Fit 4.}
\label{fig13}
\end{figure}

\begin{figure}[p]
\begin{center}
\begin{tabular}{ccc}
\includegraphics[width=8cm]{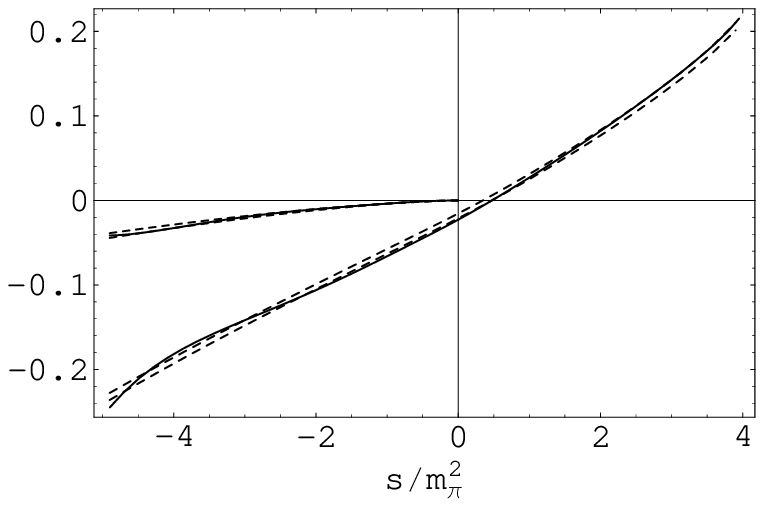}& \raisebox{-1mm}{$\includegraphics[width=8cm]{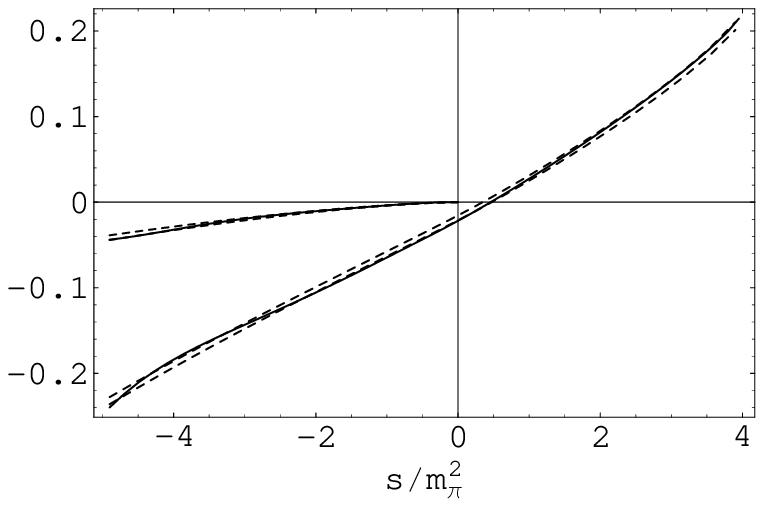}$}\\ (a)&(b)
\end{tabular}
\end{center}
\caption{The real and the imaginary parts of the amplitude $T^0_0$
of the $\pi\pi$ scattering are shown. Solid lines show our
description, dashed lines mark borders of the real part corridor
and the imaginary part for $s < 0$ from Ref. \cite{sigmaPole}: a)
Fit 3; b) Fit 4. } \label{fig14}
\end{figure}

\begin{figure}[p]
\begin{center}
\begin{tabular}{ccc}
\includegraphics[width=8cm]{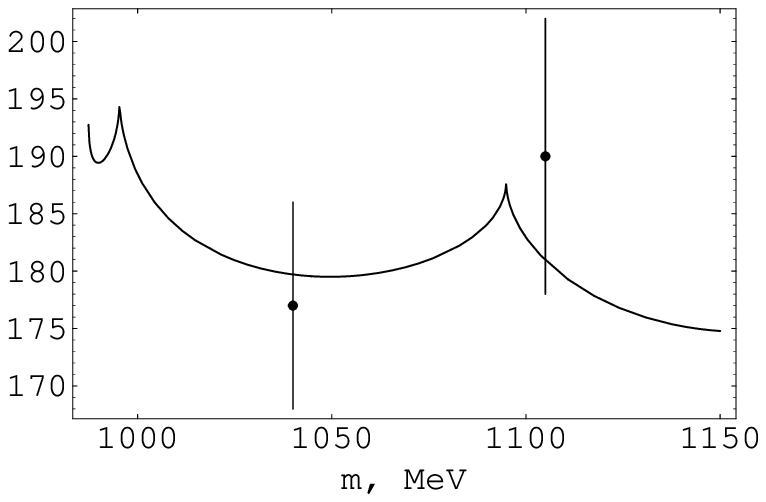}& \raisebox{-1mm}{$\includegraphics[width=8cm]{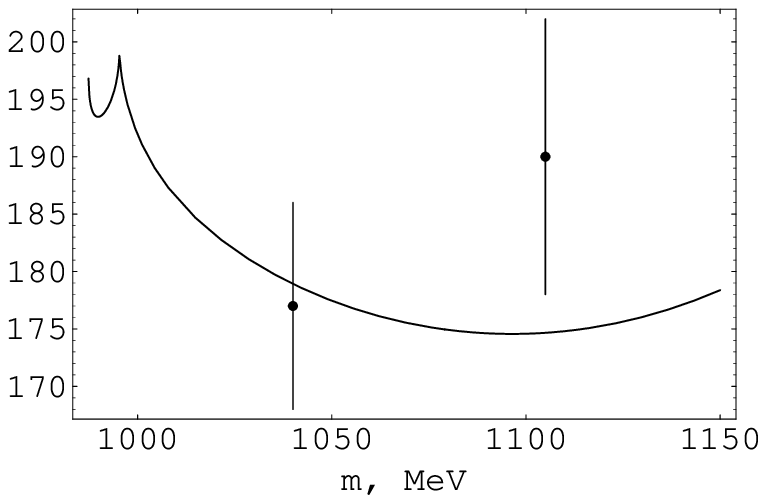}$}\\ (a)&(b)
\end{tabular}
\end{center}
\caption{The phase $\delta^{\pi K}$ of the $\pi\pi\to K\bar K $
scattering is shown: a) Fit 3; b) Fit 4.} \label{fig15}
\end{figure}

\begin{figure}[p]
\begin{center}
\begin{tabular}{ccc}
\includegraphics[width=8cm]{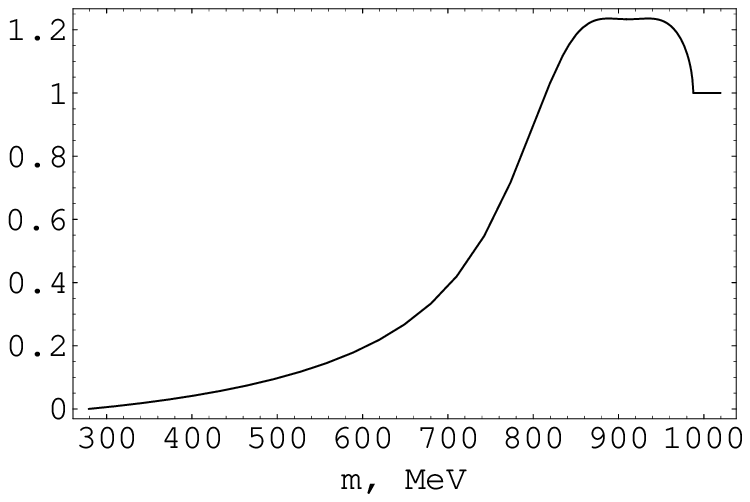}& \raisebox{-1mm}{$\includegraphics[width=8cm]{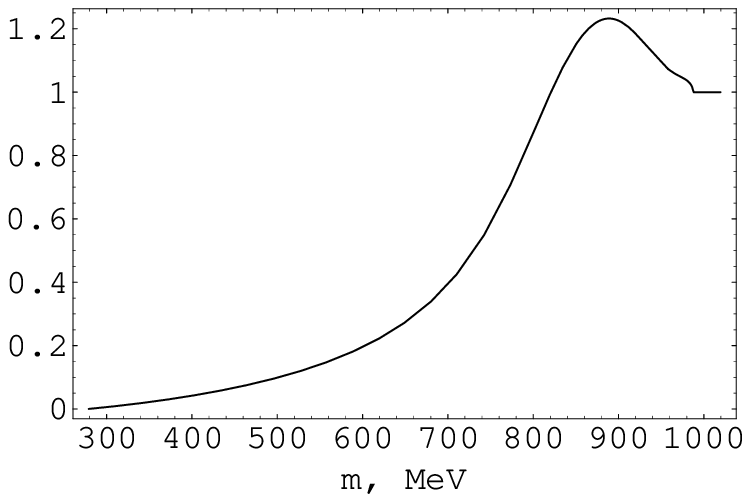}$}\\ (a)&(b)
\end{tabular}
\end{center}
\caption{The $|P_K(m)|^2$ is shown, see Eq. (7): a) Fit 3; b) Fit
4.} \label{fig16}
\end{figure}

\begin{figure}[p]
\begin{center}
\begin{tabular}{ccc}
\includegraphics[width=8cm]{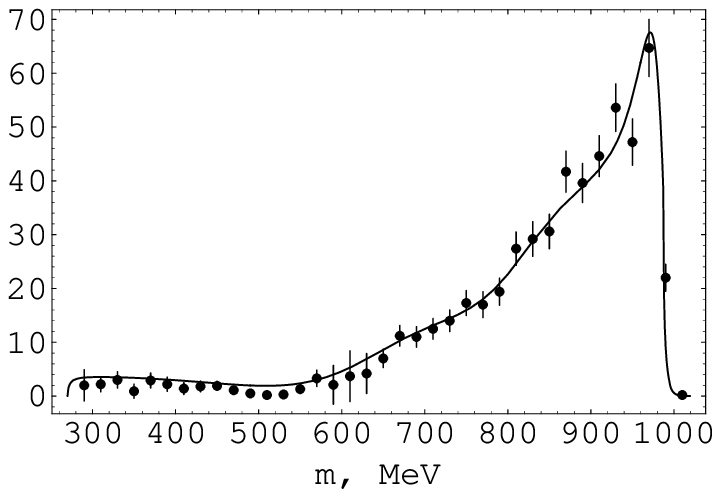}& \includegraphics[width=8cm]{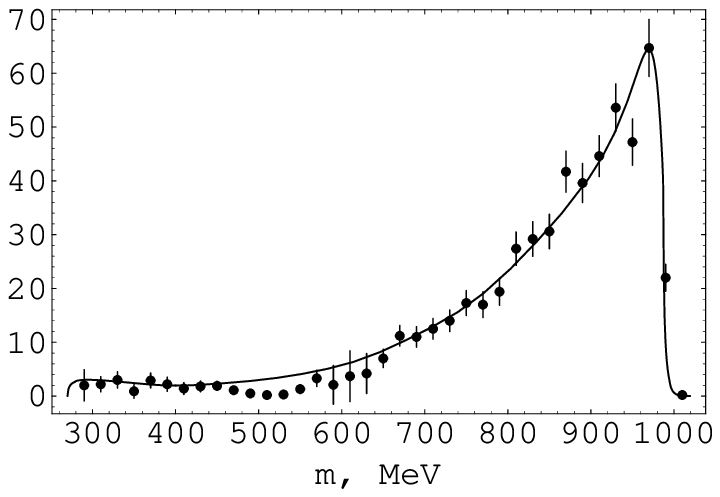}\\ (a)&(b)
\end{tabular}
\end{center}
\caption{The $\pi^0\pi^0$ spectrum in the
$\phi\to\pi^0\pi^0\gamma$ decay, theoretical curve, and the KLOE
data (points) \cite{pi0publ} are shown: a) Fit 5, b) Fit 6.}
\label{fig17}
\end{figure}

\begin{figure}[p]
\begin{center}
\begin{tabular}{ccc}
\includegraphics[width=8cm]{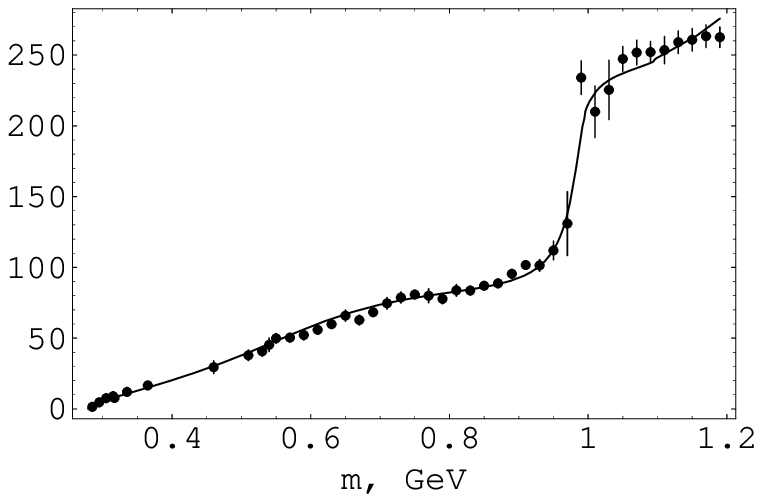}& \raisebox{-1mm}{$\includegraphics[width=8cm]{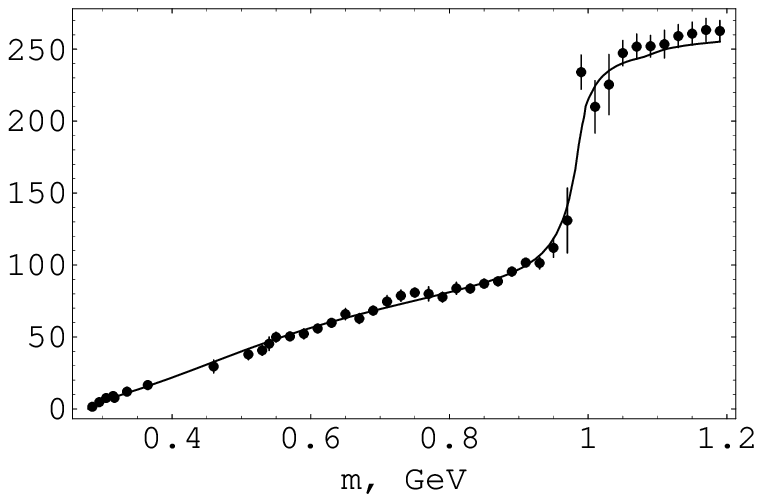}$}\\ (a)&(b)
\end{tabular}
\end{center}
\caption{The phase $\delta_0^0$ of the $\pi\pi$ scattering
(degrees) is shown: a) Fit 5, b) Fit 6. The experimental data from
Refs. \cite{hyams,estabrook,martin,srinivasan,rosselet}.}
\label{fig18}
\end{figure}

\begin{figure}[p]
\begin{center}
\begin{tabular}{ccc}
\includegraphics[width=8cm]{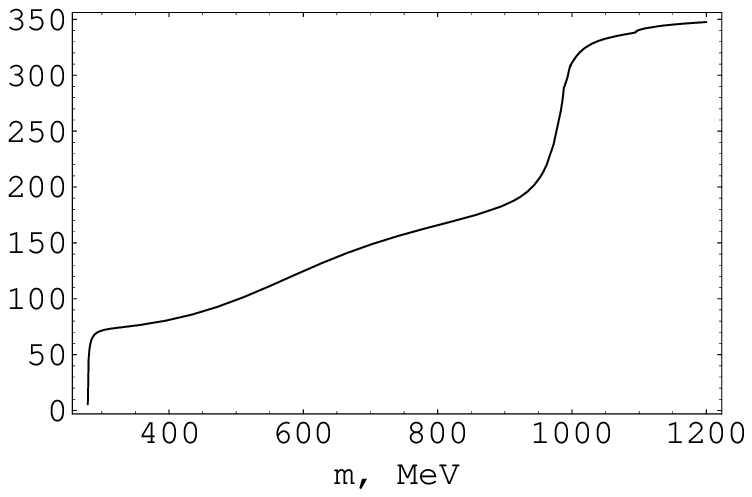}& \raisebox{-1mm}{$\includegraphics[width=8cm]{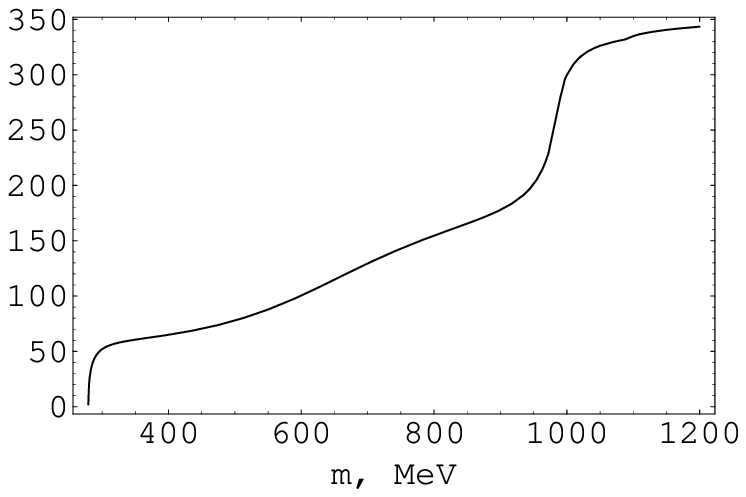}$}\\ (a)&(b)
\end{tabular}
\end{center}
\caption{The resonance phase of the $\pi\pi$ scattering
$\delta_0^{0\,res}$ (degrees) is shown: a) Fit 5, b) Fit 6.}
\label{fig19}
\end{figure}

\begin{figure}[p]
\begin{center}
\begin{tabular}{ccc}
\includegraphics[width=8cm]{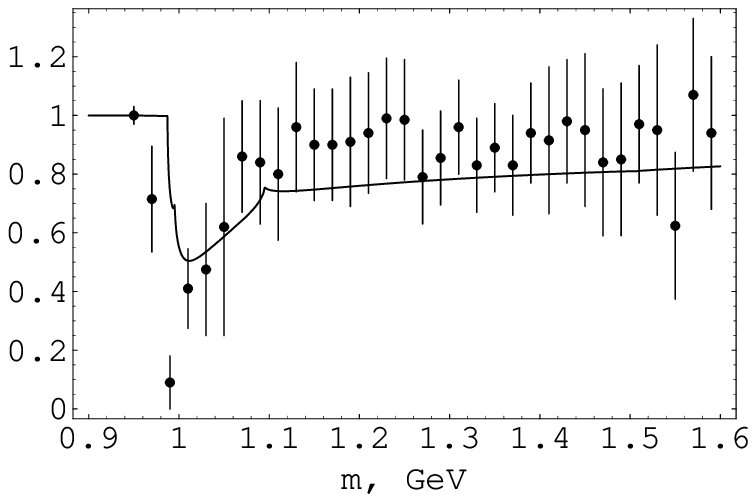}& \raisebox{-1mm}{$\includegraphics[width=8cm]{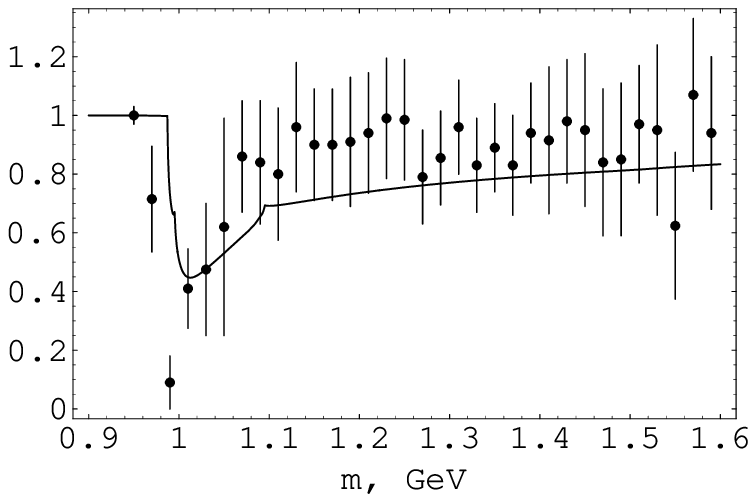}$}\\ (a)&(b)
\end{tabular}
\end{center}
\caption{The inelasticity $\eta^0_0$ is shown: a) Fit 5, b) Fit
6.} \label{fig20}
\end{figure}

\begin{figure}[h]
\begin{center}
\begin{tabular}{ccc}
\includegraphics[width=8cm]{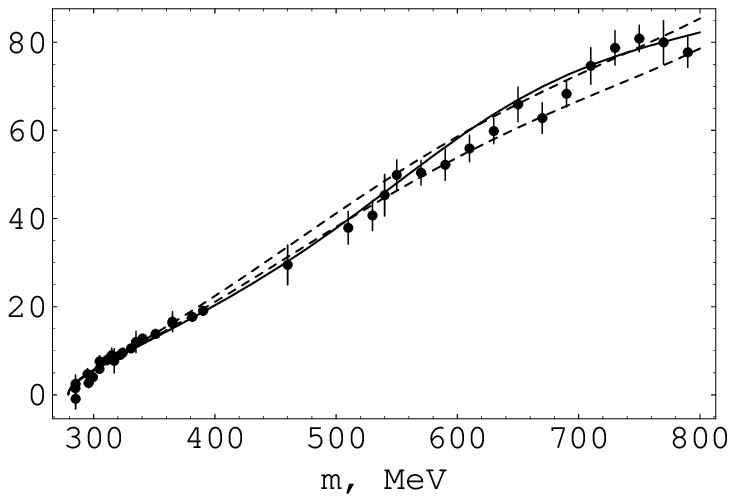}& \raisebox{-1mm}{$\includegraphics[width=8cm]{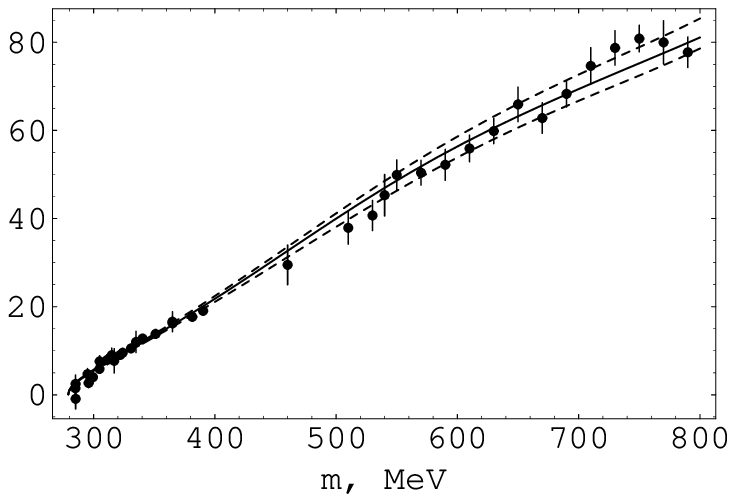}$}\\ (a)&(b)
\end{tabular}
\end{center}
\caption{The phase $\delta_0^0$ of the $\pi\pi$ scattering is
shown. The solid line is our description, dashed lines mark
borders of the corridor \cite{sigmaPole}, and points are the
experimental data from Refs. \cite{scatbnl,na48,
hyams,estabrook,martin,srinivasan,rosselet}: a) Fit 5, b) Fit 6.}
\label{fig21}
\end{figure}

\begin{figure}[h]
\begin{center}
\begin{tabular}{ccc}
\includegraphics[width=8cm]{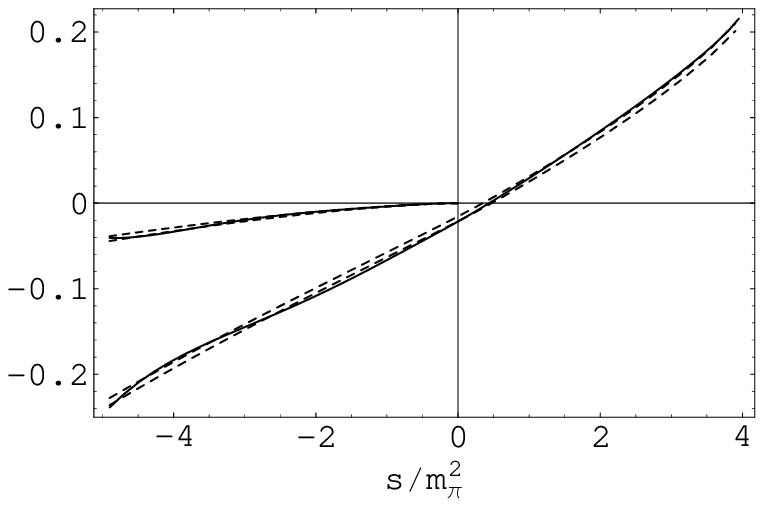}& \raisebox{-1mm}{$\includegraphics[width=8cm]{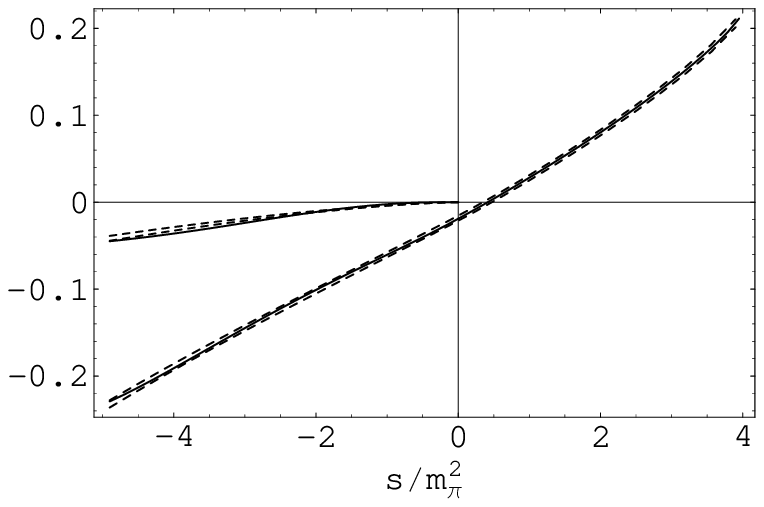}$}\\ (a)&(b)
\end{tabular}
\end{center}
\caption{The real and the imaginary parts of the amplitude $T^0_0$
of the $\pi\pi$ scattering are shown. Solid lines show our
description, dashed lines mark borders of the real part corridor
and the imaginary part for $s < 0$ from Ref. \cite{sigmaPole}: a)
Fit 5; b) Fit 6. } \label{fig22}
\end{figure}

\begin{figure}[h]
\begin{center}
\begin{tabular}{ccc}
\includegraphics[width=8cm]{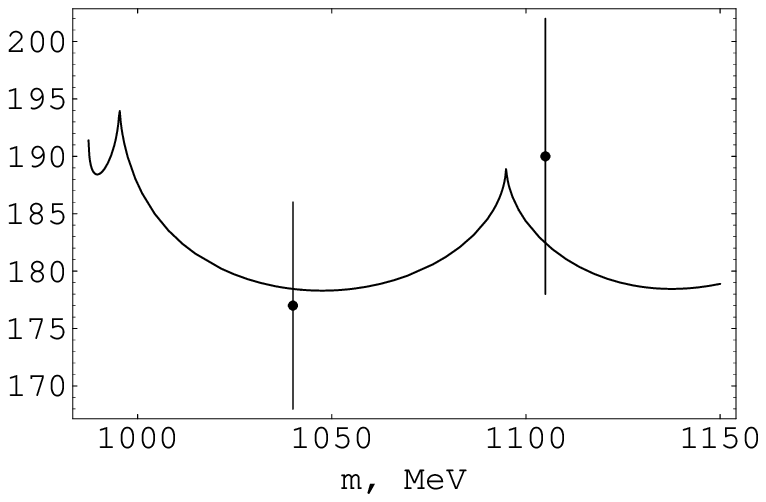}& \raisebox{-1mm}{$\includegraphics[width=8cm]{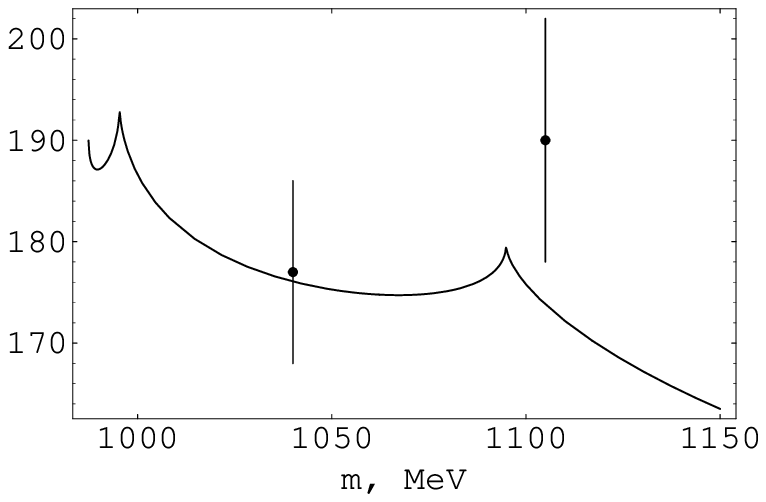}$}\\ (a)&(b)
\end{tabular}
\end{center}
\caption{The phase $\delta^{\pi K}$ of the $\pi\pi\to K\bar K $
scattering is shown: a) Fit 5; b) Fit 6.} \label{fig23}
\end{figure}

\begin{figure}[h]
\begin{center}
\begin{tabular}{ccc}
\includegraphics[width=8cm]{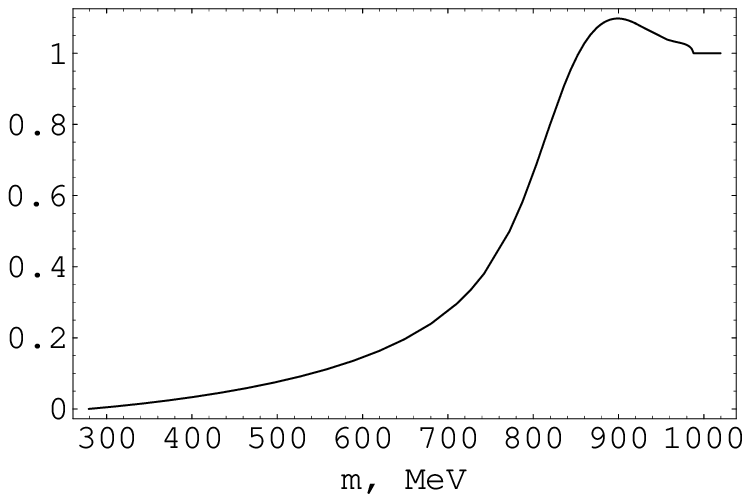}& \raisebox{-1mm}{$\includegraphics[width=8cm]{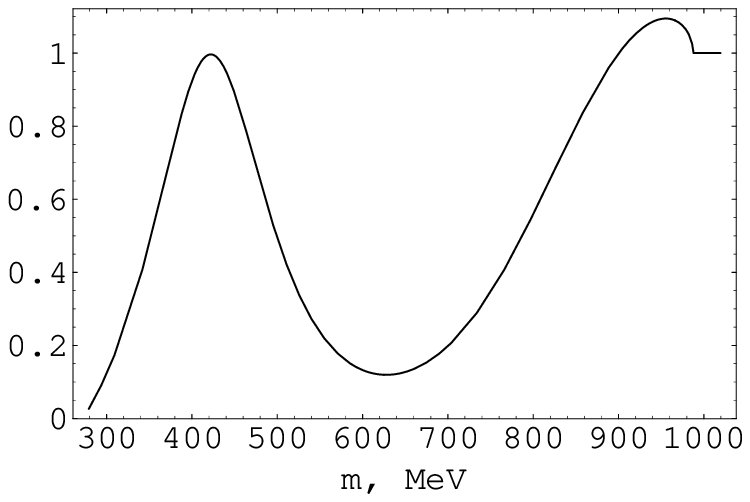}$}\\ (a)&(b)
\end{tabular}
\end{center}
\caption{The $|P_K(m)|^2$ is shown, see Eq. (7): a) Fit 5; b) Fit
6.} \label{fig24}
\end{figure}

\begin{figure}[h]
\begin{center}
\begin{tabular}{ccc}
\includegraphics[width=8cm]{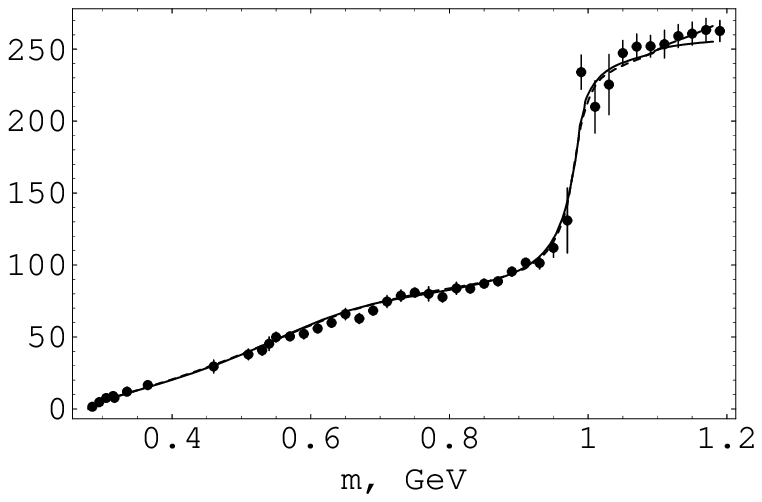}& \raisebox{-1mm}{$\includegraphics[width=8cm]{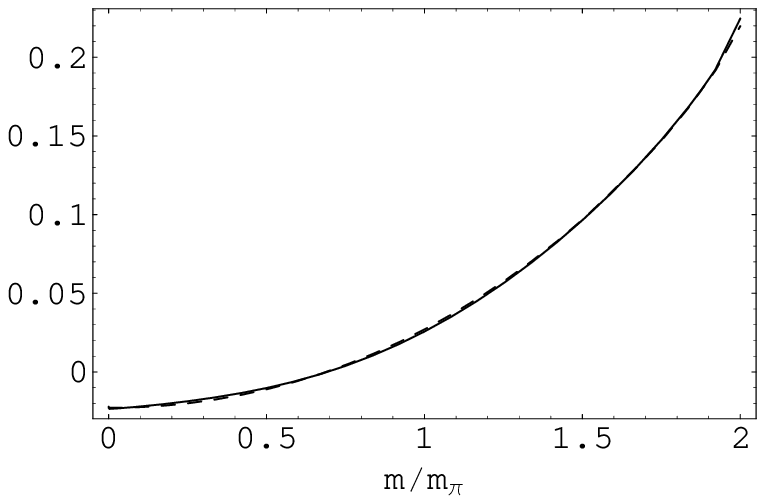}$}\\ (a)&(b)
\end{tabular}
\end{center}
\caption{The comparison of Fit 3 and Fit 7 (with the same
resonance parameters, but the background parameterization
(\ref{phB2})): a) the phase $\delta_0^0$, b) the amplitude $T^0_0$
under the $\pi\pi$ threshold. Solid lines are Fit 7, dashed lines
are Fit 3, points are the experimental data. The curves are
practically the same.} \label{fig25}
\end{figure}

\newpage
\begin{center}
\vspace{-15pt}Table VII. Parameters of the Fit 6 (with simple
background Eq. (\ref{phB2})). \vspace{2pt}
\begin{tabular}{|c|c|c|c|}\hline

$m_{f_0}$, MeV & $981.80\pm 1.8$ & $\Lambda _K$, GeV & $0.8803$
\\ \hline

$g_{f_0K^+K^-}$, GeV  & $7.3612$ & $\Lambda_1$, MeV & $490.24$ \\
\hline

$g_{f_0K^+K^-}^2/4\pi$, GeV$^2$  & $4.3120\pm 1.0$ & $\Lambda _2$,
MeV & $154.08$ \\ \hline

$g_{f_0 \pi^+\pi^-}$, GeV  & $-2.3865$ & $m_1$, MeV & $754.53$\\
\hline

$g_{f_0\pi^+ \pi^-}^2/4\pi$, GeV$^2$ & $0.453$ & $m_2$, MeV &
$422.14$ \\ \hline

$x_{f_0}$ & $0.9875$ & $w$, MeV & $0.999$ \\ \hline

$\Gamma_{f_0}(m_{f_0})$, MeV & $166.1$ & $\phi_0$ & $0.787$\\
\hline

$m_{\sigma}$, MeV & $572.25$ & $b_0$ & $1.41426$\\ \hline

$g_{\sigma\pi^+ \pi^-}$, GeV & $2.91216$ & $b_1$ & $0.97324$\\
\hline

$g_{\sigma\pi^+ \pi^-}^2/4\pi$, GeV$^2$ & $0.675$ & $b_2$ & $
-1.09477$\\ \hline

$g_{\sigma K^+K^-}$, GeV & $0.4583$ & $b_3$ & $-0.21134$
\\ \hline

$g_{\sigma K^+K^-}^2/4\pi$, GeV$^2$ & $0.017$ & $c_0$ &
$2.48601$\\ \hline

$x_\sigma$ & $1.01775$ & $c_1$ & $1.02050$\\ \hline

$\Gamma_{\sigma}(m_{\sigma})$, MeV & $387.4$ & $c_2$ & $0.45705$
\\ \hline

$C$, GeV$^2$ & $0.06582$ & $c_3$ & $0.12373$\\ \hline

$\delta $, $^{\circ}$ & $-5.8$& $\Lambda_1^{\pi}$ & $160.84$\\
\hline

$a^0_0,\ m_\pi^{-1}$ & $0.220$ & $\Lambda_2^{\pi}$ & $522.98$\\
\hline

Adler zero in $\pi\pi\to\pi\pi $ & ($89.8$ MeV)$^2$ &
$\delta_0^{0\,res}(m_\sigma)$, $^{\circ}$ & $93.1$\\ \hline

$\chi^2_{phase}$ ($44$ points) & $39.4$ &
$\delta_0^{0\,res}(m_{f_0})$, $^{\circ}$ & $251.4$\\ \hline

$\chi^2_{sp}$ ($18$ points) & $13.9$ & $\eta^0_0$($1010$ MeV) &
$0.45$\\ \hline
\end{tabular}
\end{center}

\section{Simple background}
\label{sSimpleBack}

The background function, suggested in Ref. \cite{our_f0_2011} to
reach the correct analytical properties of the $\pi\pi$ scattering
amplitude and used above, is rather complicated and costly in
computation. In this section we suggest much more simple
background parameterization, practically preserving the resonance
features, which is comfortable for experimental data analysis and
allows to describe the results \cite{sigmaPole} on the real $s$
axis.

This background function is an upgrade of the one, used in Ref.
\cite{our_f0}:

\begin{equation}
\tan(\delta^{\pi\pi}_B)=-\frac{p_\pi}{m_\pi}\frac{b_0-b_1\frac{p_\pi
^2}{m_\pi^2}+ b_2\frac{p_\pi ^4}{m_\pi^4}+b_3\frac{p_\pi
^6}{m_\pi^6}+\frac{m}{m_\pi}\bigg(c_0+c_1\frac{p_\pi ^2}{m_\pi^2}+
c_2\frac{p_\pi ^4}{m_\pi^4}+c_3\frac{p_\pi
^6}{m_\pi^6}\bigg)}{(1+4p_\pi^2 /\Lambda^{\pi\,2}_1)(1+4p_\pi^2
/\Lambda^{\pi\, 2}_2)}\,, \label{phB2}
\end{equation}

\noindent here $p_\pi=\sqrt{m^2-4m_\pi^2}/2$. Note that in
comparison with Ref. \cite{our_f0} the function (\ref{phB2}) has a
left cut.

Let us build the $\chi^2$ function. It may be devided into 3
parts: $$\chi^2=\chi^2_{data}+\chi^2_{Roy}+\chi^2_{restr}$$

\noindent where the first one is the usual $\chi^2$ function for
the experimental data, the second one provides the description of
the results \cite{sigmaPole}, and the third one provides the
restrictions.

The $\chi^2_{data}$ is constructed with the help of the same data,
as in Ref. \cite{our_f0_2011}, except the $\delta^0_0$ data in the
region $2m_\pi<m<800$ MeV, where we use the \cite{sigmaPole}
results. Note that in Table I we show $\chi^2_{phase}$, obtained
in the full region $2m_\pi<m<1200$ MeV with the "old data"
\cite{hyams,estabrook,martin,srinivasan,rosselet}.

The $\chi^2_{Roy}$ caused by the real and imaginary parts of the
$T^0_0(m)$ contributions in the region $-5m_\pi^2<s<4m_\pi^2$, and
the $\delta_0^0$ contribution from the region $4m_\pi^2 <s< (800$
MeV$)^2$. Here for $ReT^0_0$ and $\delta_0^0$ we used points and
errors, kindly sent us by H. Leutwyler, and for $ImT^0_0$ the
approximate curve $ImT^0_0(m)=-0.0327(m/2m_\pi)^3$, obtained using
Fig. 1 in Ref. \cite{sigmaPole}, providing us with central values,
and the error is assumed to be $25\%$. Note that for $ImT^0_0$ we
used the "reper" points $s=-(30$ MeV$)^2$, $-(50$ MeV$)^2$,
$-(100$ MeV$)^2$, $-(150$ MeV$)^2$, $-(200$ MeV$)^2$, $-(250$
MeV$)^2$, $-(280$ MeV$)^2$, $-(308.95$ MeV$)^2$, the last is the
end of the domain of validity of the Roy equations, connected with
the Lehmann-Martin ellipse, see \cite{sigmaPole}.

We impose the following set of restrictions, contributing to
$\chi^2_{restr}$:

\noindent 1) $85^\circ<\delta_0^{0\,res}(m_\sigma) < 95^\circ$ and
$250^\circ<\delta_0^{0\,res}(m_{f_0}) < 290^\circ$ to provide
small $\sigma-f_0$ mixing, a kind of diagonalization that results
in the four-quark model scenario.

\noindent 2) $1.2>|P_K|^2>0.8$ for $m>850$ MeV. The maximum is
found dynamically (at every calculation of the $\chi^2$ function),
the minimum in our situation is at $850$ MeV.

\noindent 3) $-0.1>\delta >-1.5$, trying to be not far from the
result \cite{rhophase}.

\noindent 4) $0.1<w<1$; $0.1$ GeV $<m_2<1.5$ GeV; $0.5$ GeV
$<\Lambda_1<2.2$ GeV; $65$ MeV $<\Lambda_2$ to provide reasonable
form of the $|P_K|^2$.

To provide, for example, the condition $\delta>-0.1$, we add to
$\chi^2_{restr}$ the term

\begin{equation}
T=W(-\delta-0.1+|\delta+0.1|)^2\,, \label{setBarrier}
\end{equation}

\noindent where $W$ is the big number. So till $\delta>-0.1$ the
contribution $T$ is equal to $0$, but when $-0.1>\delta$, $T$
becomes large, so the minimization procedure can go outside the
barrier only on a negligible distance. Our $\chi^2_{restr}$ is the
sum of contributions like Eq. (\ref{setBarrier}).

Using the constructed $\chi^2$ function, we obtain Fit 6. One can
see that this Fit perfectly describes the experimental data and
the results based on Roy equations on the real $s$ axis, see Table
VII and Figs. 17-24. Note that in Table VII the $m_{f_0}$ and
$g_{f_0K^+K^-}^2/4\pi$ errors are adduced.

To illustrate the abilities of the background (\ref{phB2}), we
perform Fit 7 with the same resonance parameters as for Fit 3. Fit
7 provides practically the same experimental data description as
Fit 3. The theoretical curves for phase $\delta^0_0$ are shown in
Fig. 25 (a), they are practically the same. It is obvious that
both Fit 3 and Fit 7 provide practically identical mass spectrum
in $\phi\to\pi^0\pi^0\gamma$ decay also. The inelasticity is
exactly the same. Additionally, Fit 7 and Fit 3 provide
indistinguishable curves for $T^0_0$ at $4m_\pi^2>s>0$, see Fig.
25 (b).

\section{Conclusion}
\label{sConclusion}

Our investigation shows that the scenario, based on the four-quark
model, completely agrees with the current experimental data and
theoretical requirements. It is shown that the requirement of the
weak $\sigma(600)-f_0(980)$ mixing leads to the $g_{\sigma
K^+K^-}$ and $g_{f_0 \pi^+\pi^-}$ suppression, that is predicted
by the four-quark model, see Table I.

The behaviour of the factor $P_K(m)$, which corrects the kaon loop
model, is model dependent. We show that for large enough $g_{f_0
K^+K^-}^2/4\pi$ constant (f.e., $1.5$ GeV$^2$) the current data
(including the Ref. \cite{sigmaPole} results) may be
well-described with this factor close to $1$ at 850 MeV $<m$ , but
for smaller values of this constant (f.e., $1$ GeV$^2$) the
correction increases. New precise data on the $\pi\pi\to K\bar K$
reaction and the inelasticity ($\eta^0_0$) of the $\pi\pi$
scattering would give an ability to understand more about this
factor and reduce the region of possible values of parameters.

The obtaining of the $\sigma$ pole in Ref. \cite{sigmaPole} gave
the strong argument in favor of the $\sigma(600)$ existence, but
followed efforts aiming the precise determination of the pole are
not productive. The Roy equations are one-channel, that is, are
approximate and even slight discrepancy in the amplitude in the
physical region may lead to large changes in the complex plane.
Remind that the Riemannian surface of the $\pi\pi$ scattering
amplitude has many sheets (strictly speaking, infinite number of
sheets), and even for relatively narrow $f_0(980)$ is sometimes a
problem to determine on what sheet should we find the pole, see,
for example, Fit 2 in Ref. \cite{our_f0_2011}. But even if we
obtained the pole precisely, it would give us practically no
information on the resonance nature, because it can not be
connected to coupling constant in the Hermitian (or
quasi-Hermitian) Hamiltonian, see also Ref. \cite{annshgn-07},
because of large imaginary part. Besides, the residue of the
amplitude in the pole is strongly distorted by the background part
of the amplitude, see Tables V and VI, that gives essential
contribution even for relatively narrow $f_0(980)$.

Let us dwell on the results, presented in Table VI. Remind that
for a stable particle with the mass $m_0$ there is the pole in the
amplitude $$T=-\frac{g^2/16\pi}{s-m_0^2} $$ \noindent at
$s=m_0^2$, the residue of the amplitude $ResT$ is connected to the
coupling constant ($g$) of the stable particle with the $\pi\pi$
channel.

One can see that the real part of the $T^0_0$ residue in the
$f_0(980)$ pole is positive, so the coupling constant should be
practically pure imaginary, what is physically meaningless. Note
that the residue of the amplitude resonance part $T^{0\,Res}_0$ is
good. That is why the best way for understanding the nature of the
light scalars is the investigation of their production mechanisms
in physical processes.

The simple background parameterization, suggested in Sec.
\ref{sSimpleBack}, may be used for experimental data analysis and
the description of the Ref. \cite{sigmaPole} results for real $s$.
It is shown that the resonance features are practically preserved,
moreover, one can see that for even more simple background, used
in Ref. \cite{our_f0}, they changed not so much, though the Ref.
\cite{sigmaPole} results
were not included.% \cite{polesOfOldBckg}.

In this investigation we paid more attention to the inelasticity
$\eta^0_0$, namely, we tried to reproduce the peculiar behaviour
near the threshold, indicated by the experimental data.
Unfortunately, the current data have large errors, so the precise
measurement of the inelasticity $\eta^0_0$ near 1 GeV in
$\pi\pi\to\pi\pi $ would be very important.

\section{Acknowledgements}
We thank very much H. Leutwyler for providing numerical values of
the $T^0_0(s)$ on the real axis, obtained in Ref.
\cite{sigmaPole}, useful discussions, and kind contacts. This work
was supported in part by RFBR, Grant No 10-02-00016.


\begin{thebibliography}{99}

\bibitem{our_f0}
N.N. Achasov and A.V. Kiselev, Phys. Rev. {\bf D73}, 054029
(2006); Erratum-ibid. {\bf D74}, 059902 (2006); Yad. Fiz. {\bf
70}, 2005 (2007) [Phys. At. Nucl. {\bf 70}, 1956 (2007)].

\bibitem{pi0publ} A.Aloisio et al. (KLOE Collaboration), Phys. Lett. {\bf
B537}, 21 (2002).

\bibitem{achasov-89}
N.N. Achasov and V.N. Ivanchenko, Nucl. Phys. {\bf B315}, 465
(1989).

\bibitem{achasov-97}
N.N. Achasov and V.V. Gubin, Phys. Rev. {\bf D56}, 4084 (1997).

\bibitem{a0f0}
N.N. Achasov, V.V. Gubin, Phys. Rev. {\bf D63}, 094007 (2001).

\bibitem{achasov-03}
N.N. Achasov, Nucl. Phys. {\bf A 728}, 425 (2003).

\bibitem{our_a0}
N.N. Achasov and A.V. Kiselev, Phys. Rev. {\bf D68}, 014006
(2003).

\bibitem{annshgn-94}
N.N. Achasov and G.N. Shestakov, Phys. Rev. {\bf D49}, 5779
(1994).

\bibitem{annshgn-07}
N.N. Achasov and G.N. Shestakov, Phys. Rev. Lett. {\bf 99}, 072001
(2007).

\bibitem{sigmaPole}
I. Caprini, G. Colangelo and H. Leutwyler, Phys. Rev. Lett. {\bf
96}, 132001 (2006).

\bibitem {our_f0_2011}
N.N. Achasov and A.V. Kiselev, Phys. Rev. {\bf D83}, 054008
(2011), or arXiv:1011.4446v2 [hep-ph].

\bibitem{jaffe}
R.L. Jaffe, Phys. Rev. {\bf D15}, 267 (1977);  {\bf 15}, 281
(1977).

\bibitem {Msig}

In the kaon loop model, $\phi\to K^+K^-\to \gamma(f_0+\sigma)$
\cite{achasov-89,achasov-97,a0f0}, the amplitude of the signal
$\phi(p)\to\gamma(f_0+\sigma)\to\pi^0(k_1)\pi^0(k_2)\gamma(q)$ is
\begin{equation}
M_{sig}=g(m)\bigg((\phi\epsilon)- \frac{(\phi q)(\epsilon
p)}{(pq)}\bigg)\,T\left(K^+K^-\to\pi^0\pi^0\right )\times 16\pi
\label{f0signal}\,,
\end{equation}
where $g(m)$ is the kaon loop function, $\phi$ and $\epsilon$ are
polarization vectors of the $\phi$ meson and photon, see Ref.
\cite{our_f0,our_f0_2011}.

\bibitem{equalConstants}
These constants are equal in the naive four-quark model.

\bibitem{hyams}
B. Hyams et al., Nucl. Phys. {\bf B64}, 134 (1973).
\bibitem{estabrook}
P. Estabrooks and A.D. Martin, Nucl. Phys. {\bf B79}, 301 (1974).
\bibitem{martin}
A.D. Martin, E.N. Ozmutlu, E.J. Squires, Nucl. Phys. {\bf B121},
514 (1977).
\bibitem{srinivasan}
V. Srinivasan et al., Phys. Rev. {\bf D12}, 681 (1975).
\bibitem{rosselet}
L. Rosselet et al., Phys. Rev. {\bf D15}, 574 (1977).

\bibitem{rhophase}
N.N. Achasov and A.A. Kozhevnikov, Phys. Rev. {\bf D61}, 054005
(2000); Yad. Fiz. {\bf 63}, 2029 (2000) [Phys. At. Nucl. {\bf 63},
1936 (2000)].

\bibitem{scatbnl}
S. Pislak et al., Phys. Rev. Lett. {\bf 87}, 221801 (2001).

\bibitem{na48}
J.R. Batley et al., Eur. Phys. J. {\bf C54}, 411 (2008).

\bibitem{ourProp}
N.N. Achasov and A.V. Kiselev, Phys. Rev. {\bf D70}, 111901 (R)
(2004).

\bibitem{sndphi}
 M.N. Achasov et al, Phys. Rev. {\bf D63}, 072002 (2001).

\bibitem{dolinsky}
S.I. Dolinsky et al., Z. Phys. {\bf C42}, 511 (1989).

\bibitem{pigam}
M.N. Achasov et al, Phys. Lett. {\bf B559}, 171 (2003).

\bibitem{nullDerivIn0}
I. Caprini, Phys.Rev. {\bf D} 77, 114019 (2008).
\bibitem{2nullDerivIn0}
 G. Colangelo, J. Gasser and H. Leutwyler, Nucl. Phys. B 603 125
(2001).

\bibitem{pkphase}
A. Etkin et al, Phys. Rev. {\bf D25}, 1786 (1982).

%\bibitem{polesOfOldBckg}
%It is insructive that the background (\ref{phB2}) has several
%poles on the physical sheet.

\end{thebibliography}
\end{document}